# CHANGES IN THE PHYSICAL ENVIRONMENT OF THE INNER COMA OF 67P/CHURYUMOV-GERASIMENKO WITH DECREASING HELIOCENTRIC DISTANCE

D. BODEWITS[1§], L. M. LARA[2], M. F. A'HEARN[1,3], F. LA FORGIA[4], A. GICQUEL[5], G. KOVACS[5], J. KNOLLENBERG[6], M. LAZZARIN[4], Z.–Y. LIN (林忠義 )[7], X. SHI[5], C. SNODGRASS[8], C. TUBIANA[5], H. SIERKS[5], C. BARBIERI[4], P. L. LAMY[9], R. RODRIGO[10,11], D. KOSCHNY[12], H. RICKMAN[13], H. U. KELLER[14], M. A. BARUCCI[15], J.–L. BERTAUX[16], I. BERTINI[17], S. BOUDREAULT[5], G. CREMONESE[18], V. DA DEPPO[19], B. DAVIDSSON[13], S. DEBEI[20], M. DE CECCO[21], S. FORNASIER[15], M. FULLE[22], O. GROUSSIN[9], P.J. GUTIÉRREZ[2], C. GÜTTLER[5], S. F. HVIID[6], W.-H. IP[23], L. JORDA[9], J.-R. KRAMM[5], E. KÜHRT[6], M. KÜPPERS[24], J. J. LÓPEZ-MORENO[2], F. MARZARI[4], G. NALETTO[25,4,19], N. OKLAY[5], N. THOMAS[26, 27], I. TOTH[28], J.-B. VINCENT[5]

1. Department of Astronomy, University of Maryland, College Park, MD 20742-2421, USA
§ Corresponding author. Email: Dennis@astro.umd.edu
2. Instituto de Astrofisica de Andalucia-CSIC, Glorieta de la Astronomia, 18008 Granada, Spain
3. Gauss Professor, Akademie der Wissenschaften zu Göttingen, 37077 Göttingen, Germany
4. University of Padova, Department of Physics and Astronomy, Vicolo dell'Osservatorio 3, 35122, Italy
5. Max-Planck Institut für Sonnensystemforschung, Justus-von-Liebig-Weg, 3 37077 Göttingen, Germany
6. Deutsches Zentrum für Luft- und Raumfahrt (DLR), Institut für Planetenforschung, Rutherfordstrasse 2, 12489 Berlin, Germany
7. Institute of Astronomy, National Central University, Chung-Li 32054, Taiwan
8. Planetary and Space Sciences, Department of Physical Sciences, The Open University, Milton Keynes, MK7 6AA, UK
9. Aix Marseille Université, CNRS, LAM (Laboratoire d'Astrophysique de Marseille), UMR 7326, 13388 Marseille, France
10. Centro de Astrobiología, CSIC-INTA, 28850 Torrejón de Ardoz, Madrid, Spain
11. International Space Science Institute, Hallerstraße 6, 3012 Bern, Switzerland
12. Scientific Support Office, European Space Research and Technology Centre/ESA, Keplerlaan 1, Postbus 299, 2201 AZ Noordwijk ZH, The Netherlands
13. Department of Physics and Astronomy, Uppsala University, Box 516, 75120 Uppsala, Sweden
14. Institut für Geophysik und extraterrestrische Physik (IGEP), Technische Universität Braunschweig, Mendelssohnstr. 3, 38106 Braunschweig, Germany
15. LESIA-Observatoire de Paris, CNRS, Universite Pierre et Marie Curie, Universite Paris Diderot, 5, Place J. Janssen, 92195 Meudon Principal Cedex, France
16. LATMOS, CNRS/UVSQ/IPSL, 11 Boulevard d'Alembert, 78280 Guyancourt, France
17. Centro di Ateneo di Studi ed Attivitá Spaziali "Giuseppe Colombo" (CISAS), University of Padova, Via Venezia 15, 35131 Padova, Italy
18. INAF, Osservatorio Astronomico di Padova, Vicolo dell'Osservatorio 5, 35122 Padova, Italy
19. CNR-IFN UOS Padova LUXOR, Via Trasea, 7, 35131 Padova, Italy
20. Department of Industrial Engineering, University of Padova, Via Venezia, 1, 35131 Padova, Italy
21. University of Trento, via Sommarive, 9, 38123 Trento, Italy
22. INAF - Osservatorio Astronomico di Trieste, Via Tiepolo 11, 34014 Trieste, Italy
23. Institute for Space Science, National Central University, 32054 Chung-Li, Taiwan
24. Operations Department, European Space Astronomy Centre/ESA, P.O. Box 78, 28691 Villanueva de la Cañada (Madrid), Spain
25. Department of Information Engineering, University of Padova, Via Gradenigo 6/B, 35131 Padova
26. Physikalisches Institut der Universität Bern, Sidlerstr. 5, 3012 Bern, Switzerland
27. Center for Space and Habitability, University of Bern, 3012 Bern, Switzerland
28. MTA CSFK Konkoly Observatory, Konkoly Thege M. ut 15/17, HU 1525 Budapest, Hungary

## ABSTRACT

The Wide Angle Camera of the OSIRIS instrument on board the *Rosetta* spacecraft is equipped with several narrowband filters that are centered on the emission lines and bands of various fragment species. These are used to determine the evolution of the production and spatial distribution of the gas in the inner coma of comet 67P with time and heliocentric distance, here between 2.6 – 1.3 AU pre-perihelion. Our observations indicate that the emission observed in the OH, OI, CN, NH, and $NH_2$ filters is mostly produced by dissociative electron impact excitation of different parent species. We conclude that $CO_2$ rather than $H_2O$ is a significant source of the [OI] 630 nm emission. A strong plume-like feature observed in the in CN and [OI] filters is present throughout our observations. This plume is not present in OH emission and indicates a local enhancement of the $CO_2/H_2O$ ratio by as much as a factor of 3. We observed a sudden decrease in intensity levels after March 2015, which we attribute to decreased electron temperatures in the first kilometers above the nucleus surface.

6 figures, 6 tables

# 1. INTRODUCTION

Comets are considered relatively pristine leftovers from the early days of our Solar System. They are distinguished from other minor bodies by a coma of gas and dust produced when ices retained from the formation of the solar system sublime. Understanding the connection between the coma and the comet's nucleus is critical because observations rarely detect the nucleus directly, and its properties must often be inferred from measurements of the surrounding coma. Because measurements of the coma do not necessarily represent the characteristics of the nucleus due to spatial, temporal and chemical evolution of the emitted material, projecting the coma observations back to the nucleus requires an understanding of the processes that induce changes in the coma. Compositional studies must take into account chemical reactions and photolysis to determine how the molecular abundances measured in the coma relate to the bulk composition of the nucleus.

Direct, high-resolution observations of cometary nuclei are rare, coming only from spacecraft encounters. Connecting these measurements to the coma provides a valuable means of evaluating the techniques used in situations where the nucleus cannot be seen. Images obtained by *Giotto, Vega, Deep Space 1* and *Stardust* showed details of their targets' nuclei (Keller et al. 1987; Soderblom et al. 2002; Brownlee et al. 2004; Veverka et al. 2013), but in each case, coma observations were limited to the continuum around closest approach. The *Deep Impact* spacecraft was the first to observe both the gas and dust comae of comets 9P/Tempel 1 and 103P/Hartley 2 through a multi-wavelength filter set, while monitoring each comet for months around close approach (A'Hearn et al. 2005; A'Hearn et al. 2011).

Orbiting the comet since August 2014, the European Space Agency's *Rosetta* mission has allowed an unprecedented study of the activity and evolution 67P/Churyumov-Gerasimenko. The large heliocentric distance of the comet, its low activity levels, and the close proximity of the spacecraft to its surface allow us to sample an environment that has never been studied before and that is not accessible for observations from Earth. This paper describes observations of the dust and gas in the coma of 67P/Churyumov-Gerasimenko acquired by *Rosetta*'s Optical, Spectroscopic, and Infrared Remote Imaging System (OSIRIS). Fragment species are relatively bright and emit in wavelengths accessible from the ground. OSIRIS' narrow band filters provide an important link to ground based observations, and help to connect our detailed knowledge of 67P/Churyumov-Gerasimenko to the wider population of comets.

# 2. OBSERVATIONS

OSIRIS consists of two bore-sighted cameras: a narrow angle camera (NAC, field of view 2.2 x 2.2 degrees) and a wide angle camera (WAC, field of view 11.4 x 12.11 degrees) (Keller et al. 2007). The WAC is best equipped to study the coma; its twelve narrowband and two medium width filters allow color discrimination and the imaging of emission lines and bands from gas and continuum in optical wavelengths (250–750 nm). We typically monitored gas and dust activity with the WAC, about once every two weeks for heliocentric distances greater than 2 AU, and once per week between 2 and 1.3 AU pre-perihelion. The nominal sequence had a set of observations once per hour for a full comet rotation (~12h25m during the observations described here; Keller et al. 2015). As both Rosetta and 67P were coming closer to Earth, the data volume available increased, allowing us to increase the observing cadence to one set of observations every 20 min, for 14h (to avoid aliasing with the comet's rotation period). Exposure times are optimized to achieve good signal-to-noise in the coma, often resulting in saturation on the nucleus and in the apparition of internal reflection artifacts (ghosts) appearing on the CCD images as a circular 'blob' to the right of the nucleus (Keller et al. 2007). The acquired images are scaled with binning (typically 4 x 4 for coma observations) and observing geometry (distance to the comet, distance to the Sun). To establish the connection with the nucleus, short-exposure, 2x2-binned images were acquired with the 375 nm and 610 nm filters along with the coma images. For this paper, we limit ourselves to data acquired during four periods when *Rosetta* conducted the so-called 'Volatile Activity Campaigns'. These campaigns were multi-instrument observations specifically designed to study the gas in the coma of 67P (including OSIRIS, VIRTIS, MIRO, and Alice), and two more dedicated OSIRIS campaigns to provide better temporal coverage. The observations discussed in this manuscript were acquired before the comet's perihelion, between 2015 January 24 and July 3 (see Table 1). During this period, the comet's distance to the Sun decreased from 2.48 AU to 1.34 AU. Because *Rosetta* was close to the surface in January, the nucleus fills a significant part of the field of view, whereas during the later observations the WAC maps a much larger part of the coma. For the analysis described below, we used images acquired at the approximate same diurnal phase.

**Table 1 –** Observing log.

| SEQUENCE | Date (UTC) | Time (UTC) | $r_h$ (AU) | $\Delta r_h$ (km/s) | Range (km) | Phase (deg) |
|---|---|---|---|---|---|---|
| **MTP12/DEEP VAC** | 2015 Jan 24 | 11:35 AM | 2.47 | -12.8 | 27.9 | 92.8 |
| **MTP14/DEEP VAC** | 2015 Mar 12, | 1:27 AM | 2.12 | -13.3 | 80.7 | 50.6 |
| **MTP15/STP051** | 2015 Apr 14 | 4:25 PM | 1.88 | -13.2 | 170 | 74.5 |
| **MTP16/STP055** | 2015 May 12 | 2:11 AM | 1.66 | -12.5 | 155 | 71.6 |
| **MTP17/DEEP VAC** | 2015 June 3 | 8:04 AM | 1.51 | -10.9 | 232 | 89.3 |
| **MTP18/VAC** | 2015 July 3 | 8:19 AM | 1.34 | -7.77 | 168 | 89.7 |

## 3. ANALYSIS

### *3.1 Data Reduction and Image Processing*

All images are pre-processed using the standard OSIRIS pipeline (Tubiana et al. 2015), which includes bias and dark subtraction, flat fielding, conversion from electron yield to radiance units (W $m^{-2}$ $sr^{-1}$ $nm^{-1}$), and bad-pixel masking.

#### *3.1.1. Coma gas emissions in the filters*

The narrowband filters were designed to sample either emission lines and bands of specific gases or continuum light at nearby wavelengths, but inevitably also sample the emission of other molecules with lines that fall within the narrowband filters' passbands at various levels. A summary of the characteristics of the WAC's filters and of the most prominent emission features within their pass bands is given in Table 2.

**Table 2 –** Characteristics of the WAC's filters (Keller et al. 2007).

| Filter + Name | $\lambda_{central}$ (nm) | Width (nm) | Comments |
|---|---|---|---|
| **21 Green** | 537.2 | 63.2 | Includes $C_2$ Swan $\Delta\nu$ = -1,0. |
| **31 UV245** | 246.2 | 14.1 | Isolated pinholes. |
| **41 CS** | 259.0 | 5.6 | Several overlapping pinholes. Samples CS $A^1\Pi - X^1\Sigma^+$ (0-0). |
| **51 UV295** | 295.9 | 10.9 | |
| **61 OH** | 309.7 | 4.1 | Several pinholes, one strong (at the lower left quadrant of the CCD images). Samples OH $A^2\Sigma^+ - X^2\Pi^i$ (0-0). |
| **71 UV325** | 325.8 | 10.7 | Many overlapping pinholes. |
| **81 NH** | 335.9 | 4.1 | Samples NH $A^3\Pi_1 - X^3\Sigma^-$ (0-0). Includes $OH^+$ ($A^3\Pi$ – $X^3\Sigma^-$). |
| **12 Red** | 629.8 | 156.8 | Broad band filter. |
| **13 UV375** | 375.6 | 9.8 | Minor contribution $C_3$ Comet Head Group. |
| **14 CN** | 388.4 | 5.2 | Samples CN violet system, minor contribution $C_3$ Comet Head Group. Includes $CO_2^+$ ($\tilde{A}\,^2\Pi$ – $X^3\Sigma^-$). |
| **15 NH2** | 572.1 | 11.5 | Samples $NH_2$ $\tilde{A}\,^2A_1 \rightarrow X\,^2B_1$ (0,10,0). |
| **16 Na** | 590.7 | 4.7 | Covers both Na D1 and D2 doublets; includes $C_2$ Swan $\Delta\nu$ = -2 |
| **17 OI** | 631.6 | 4.0 | Samples [OI] $^1D$ – $^3P$ 630 nm line only. Includes $NH_2$ $\tilde{A}\,^2A_1 \rightarrow X\,^2B_1$ (0,8,0). |
| **18 Vis610** | 612.6 | 9.8 | Includes $NH_2$ $\tilde{A}\,^2A_1 \rightarrow X\,^2B_1$ (0,9,0). |

The WAC can map the distribution of water with its OI and OH filters. The OI filter covers the forbidden transitions from the OI ($2p^4$) $^1D$ state to the ground state. The OI ($2p^4$) $^1D$ state is populated directly by photodissociation of $H_2O$ molecules, as is the OI ($2p^4$) $^1S$ state, which relaxes mostly (95%) by decay into the $^1D$ state (c.f. Cochran 2008). The OH filter covers the (0-0) band of the A $^2\Sigma^+$ – X $^2\Pi$ transition of OH, centered at about 308.5 nm, which is excited almost entirely by fluorescence of sun light (c.f. Schleicher & A'Hearn 1988). A small fraction of the photodissociation of $H_2O$ also leads directly to the population of OH in high rotational states of the A $^2\Sigma^+$ electronic state (A'Hearn et al. 2015), but the resulting emission falls outside the passband of the WAC's OH filter. The WAC's CN filter covers emission from $B^2\Sigma^+$ – $X^2\Sigma^+$ (0,0) transitions around 388 nm. The $NH_2$ $\tilde{A}\,^2A_1 \rightarrow X\,^2B_1$ (0,10,0) band is very wide and how much of the emission falls within the $NH_2$ filter's passband depends on the heliocentric distance and velocity. The NH filter covers the NH $A^3\Pi_1 - X^3\Sigma^-$ (0-0) transition. While the WAC is equipped with CS and Na filters, those two filters were used only sporadically during the first half of the mission because the low SNR and pinholes (CS) and contamination by $C_2$ emission (Na) hampers the interpretation of observations made with these two filters.

**Table 3 –** Continuum removal factors $F$ for the WAC gas filters assuming different continuum reddening (% per 100 nm between 375 nm and 610 nm). $\Delta F$ is the relative, error-propagated 1-sigma standard deviations of the removal factors.

| FILTERS | Removal Factor | | | | ΔF |
|---|---|---|---|---|---|
| | 0% | 10% | 20% | 30% | (%) |
| **OI/UV375** | 1.570 | 2.026 | 2.623 | 3.438 | 5 |
| **CN/UV375** | 1.021 | 1.034 | 1.053 | 1.077 | 6 |
| **OH/UV375** | 0.479 | 0.444 | 0.397 | 0.332 | 10 |
| **NH/UV375** | 0.829 | 0.790 | 0.740 | 0.671 | 8 |
| **NH2/UV375** | 1.738 | 2.124 | 2.630 | 3.321 | 5 |
| **Na/UV375** | 1.673 | 2.080 | 2.614 | 3.343 | 8 |
| **Vis610/UV375** | 1.648 | 2.127 | 2.753 | 3.609 | 5 |

The WAC is also equipped with several filters that can sample the continuum. Comparing the 375 nm and 610 nm narrowband filters illustrates the level of contamination of the latter filter by gas emission features. In WAC observations of 16Cyg, a system of two solar analogs, we measured a flux ratio Vis610/UV375 = 1.648 ± 0.08, whereas for the dust surrounding 67P this ratio was between 2.8 and 5.6 (Jan/March) and 2.5 – 3.3 (June/July). Such values would require a reddening between 20-40% per 100 nm between the two filters, while we determined from the NAC images that the average reddening of the dust was between 375 and 610 nm was around 18% per 100 nm (Sec. 3.1.2.). This suggest that in the January/March data, up to 50% of the flux in the 610 nm filter might come from gaseous emission, probably mostly due to the emission from the $NH_2$ $\tilde{A}\ ^2A_1 \rightarrow X\ ^2B_1$ (0,9,0) and $C_2$ $d^3\Pi_g - a^3\Pi_u$ ($\Delta v = -2$) transitions. As will be discussed below, the Vis610/UV375 ratio measured in June and July is close to what is expected based on the NAC reddening measurement, suggesting that the relative contribution of gaseous emission to the flux measured in the Vis610 decreased.

To further assess the gas contamination in other filters, we show the transmission of the WAC filters on a high-resolution spectrum of comet 122P/DeVico; Cochran 2002); Fig 1a). The spectrum of this comet is not a particularly good proxy for that of 67P; the comet was observed very close to the Sun (0.7 AU), was extremely dust-poor, and the two comets have different compositions (67P is depleted in its carbon chain molecules while 122P has a 'typical' composition; Fink 2009). However, the high-quality spectrum and line catalogue demonstrate the extent of the contamination and the identification of the gases responsible for it. An example is shown in Fig 1b, where we show the transmission of the OI filter overlaid on comet 122P/DeVico's spectrum. The filter was designed to sample emission from the [OI] $^1D \rightarrow {}^3P$ line at 630 nm, but also contains several emission lines of the $NH_2$ molecule, notably the $\tilde{A}\ ^2A_1 \rightarrow X\ ^2B_1$ (0,8,0). The WAC is equipped with a narrow band filter designed specifically to observe the $NH_2$ $\tilde{A}\ ^2A_1 \rightarrow X\ ^2B_1$ (0,10,0) band, which can be used to remove $NH_2$ emission from the 610 nm ($NH_2$ Ã $^2A_1$ → X $^2B_1$ (0,9,0) transition) and OI narrow band filters ($NH_2$ Ã $^2A_1$ → X $^2B_1$ transition's (0,8,0)) if the relation between the three different emission bands of $NH_2$ is known. To calculate the contribution of $NH_2$ emission to the flux measured in the $NH_2$, Vis610, and OI filters we weighted the archival spectrum *S(λ)* of 122P with OSIRIS' CCD quantum efficiency *Q(λ)*, the reflectivity *R(λ)*, and transmission of its filters *T(λ)*:

$$S'(\lambda) = \frac{\int Q(\lambda)\cdot R(\lambda)\cdot T(\lambda)\cdot S(\lambda)\cdot d\lambda}{\int Q(\lambda)\cdot R(\lambda)\cdot T(\lambda)\cdot d\lambda} \quad [1]$$

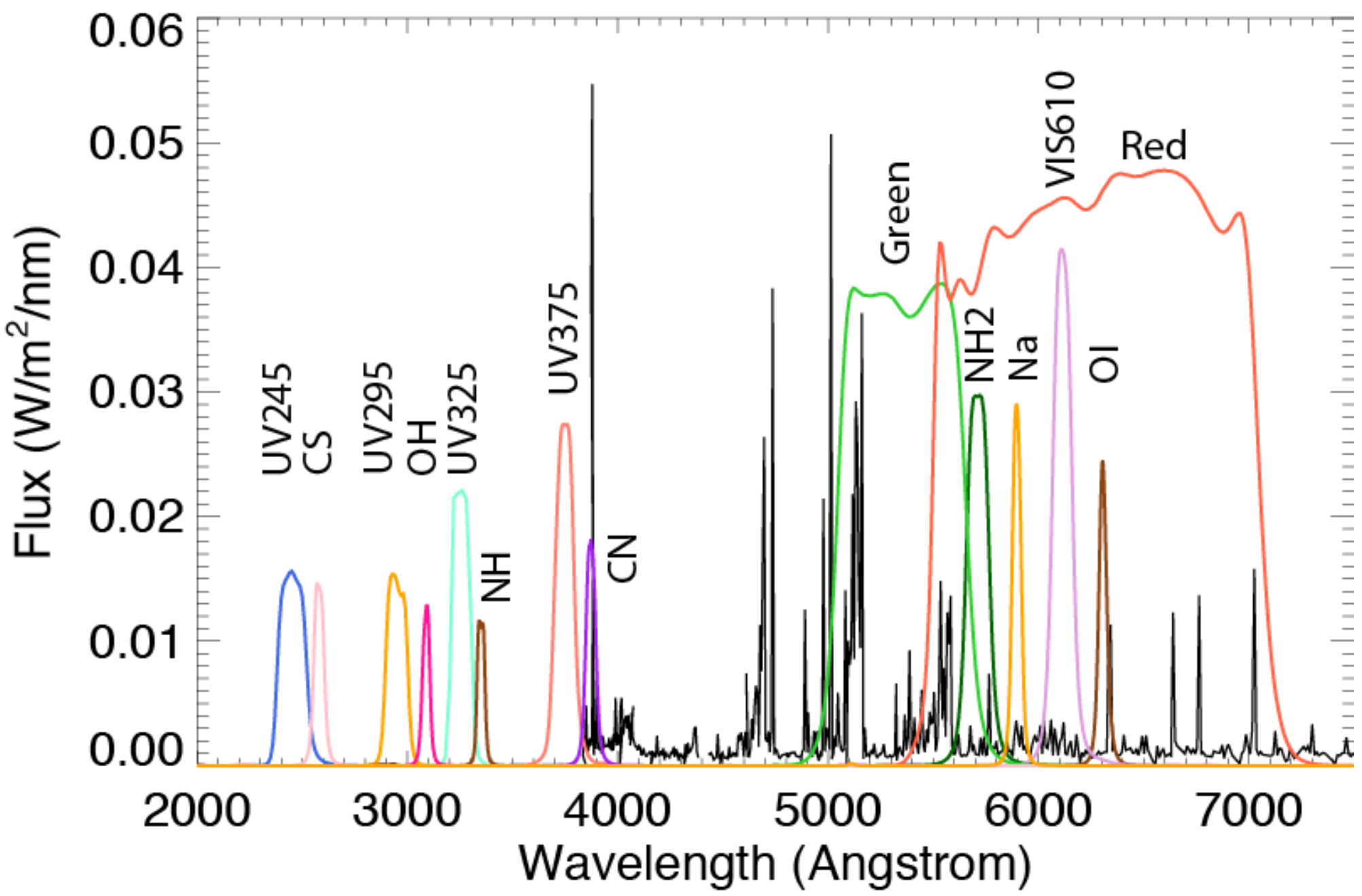

**Fig. 1a –** OSIRIS/WAC filter transmission profiles overlaid on a high-resolution spectrum of comet 122P/DeVico (Cochran et al. 2002). Profiles are not convolved with the quantum efficiency of the detector.

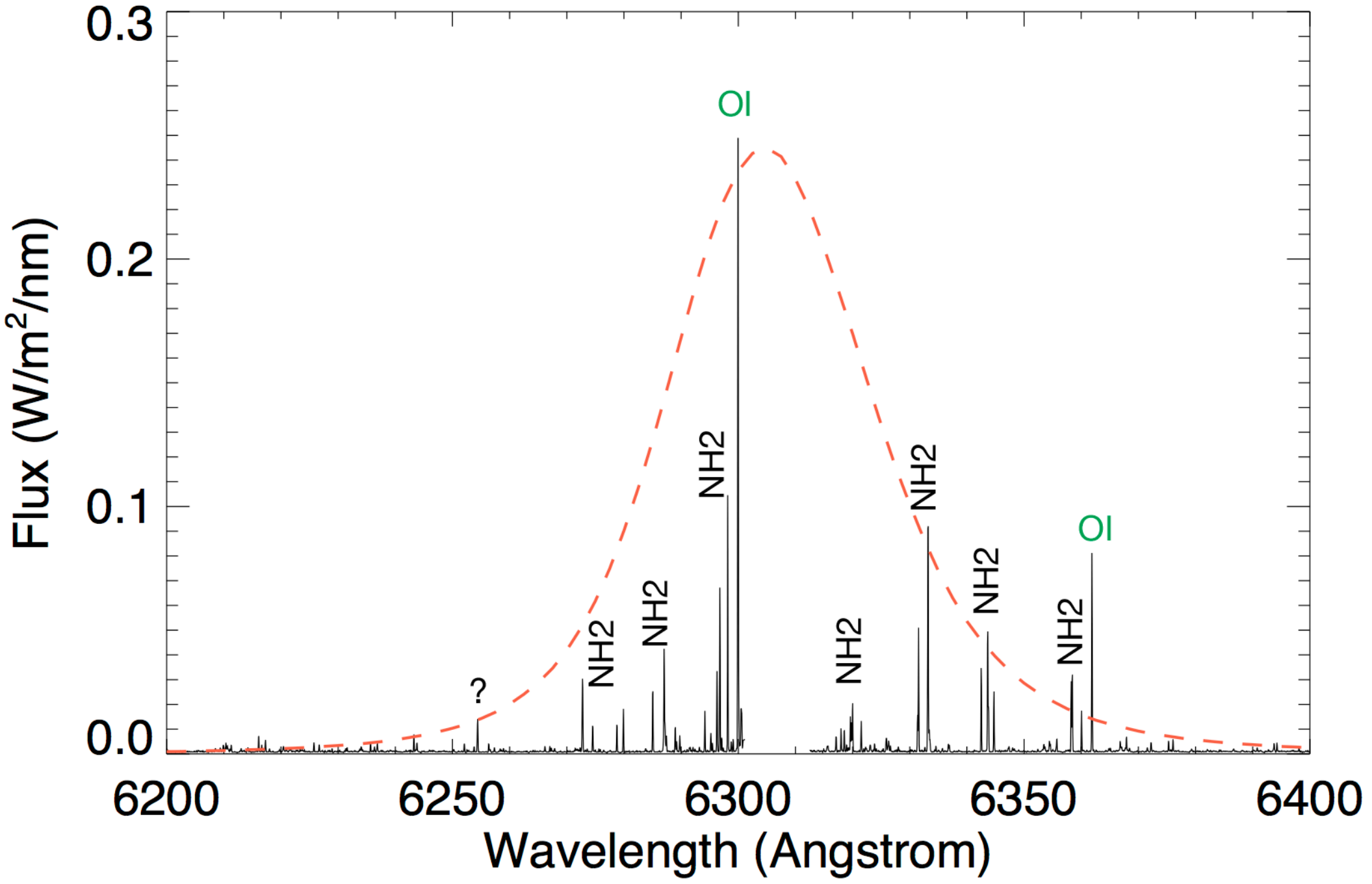

**Fig. 1b –** As above - close up on the OI filter transmission.

For comet DeVico we find that the ratio between $NH_2$ emission in the OI and $NH_2$ filters is 1.389, and 0.45 for the ratio between $NH_2$ emission in the Vis610 and $NH_2$ filters. Using this band ratio, we find that $NH_2$ typically contributed between 10-20% to the flux measured in the OI filter (after continuum subtraction); however, as we will discuss in Sec. 5, this assumes that band ratios between photo fluorescent excitation and electron impact excitation are similar, which may not be the case.

Fluorescence efficiencies for $NH_2$ have been calculated by Kawakita & Watanabe (2002). Their model shows that the ratio between fluorescence efficiencies of even and odd transitions is strongly dependent on the heliocentric distance and heliocentric velocity, but that the ratio between even bands is constant. $NH_2$ emission in the $NH_2$ and OI filters is thus coupled and can be removed with a constant factor that is independent of heliocentric distance and velocity, and this factor should be the same for both 122P and 67P. The Vis610 filter however includes emission from an odd band, (0,9,0), which has a much more complex relation to the emission in the even (0,10,0) band sampled by the $NH_2$ filter. We therefore did not attempt to remove the contribution of $NH_2$ to the images acquired with the Vis610 filter.

Concluding, we deem the filter centered at 375 nm best suited to image the continuum light, although its transmission includes some emission lines from $C_3$ (Ã $^1\Pi_u$ → $X^1\Sigma_g^+$) transitions. There are fewer emission lines at shorter wavelengths, but the three near-UV continuum filters centered on 245, 295, and 325 nm suffer heavily from pinholes (circular defects in the filters' coatings from impurities in the production process) which represent an increased light flux at shorter and longer wavelengths than those in the filter width bandpass. The disadvantage of using the UV375 filter to remove the continuum from all the gas filters is that since we use only one filter, we do not measure coma colors, and have to assume a linear reflectance with respect to the wavelength. The assumed reddening has a large impact on the continuum removal factor used with increasing wavelength difference.

### *3.1.2. Continuum removal and reddening*

To measure the gaseous emission line flux, the contribution of the reflected continuum needs to be removed (see e.g. Farnham et al. 2000). We used OSIRIS observations of the solar analogues of the 16Cyg system (the two stars are not spatially resolved by the WAC) to determine the ratio of narrowband continuum fluxes in all filters for a solar spectrum. These ratios $\alpha$ are the weight factors needed to remove the flux contribution of 'grey' dust from the total flux $F_{tot}$ measured in a narrow band using the flux $F_{375}$ measured in the UV375 filter in order to measure the flux contribution by the gas of interest:

$$F_{gas} = F_{tot} - \alpha \cdot F_{375} \quad [2]$$

The signal-to-noise ratio in the OH and NH filters is poor because UV fluxes from solar analogs are low and the CCD's quantum efficiency drops significantly below 400 nm (Magrin et al. 2015). In addition, the long exposures required in these filters increases the chance of cosmic rays affecting the measurement, reducing the number of useable images. Average filter ratios and 1-σ standard deviations are given in Table 2.

To calculate the effect of spectral reddening $P$ (in percent per 100 nm) on the flux ratios we first calculated the resulting, reddened spectrum by weighting a solar spectrum with the reddening $R(\lambda)$ as a linear function of wavelength $\lambda$ between $\lambda_1$ = 375.6 nm and $\lambda_2$ = 612.6 nm (the central wavelengths of two of the WAC's continuum filters), normalized at 494 nm:

$$R(\lambda) = 1 + 0.01 \cdot \mathrm{P} \cdot (\lambda - 0.5 \cdot (\lambda_2 - \lambda_1)) \quad [3]$$

We used the ratios of the fluxes of the reddened spectrum and the original solar spectrum to calculate continuum removal factors for increasing reddening, which are given in Table 3. To estimate the average reddening of the continuum of 67P we performed a deeper investigation of the colors of the coma using NAC medium-band filters observations, for which the contributions of extraneous emission lines should be less important. For this, we choose a dataset that had good coverage of the coma around the nucleus using a large number filters. The selected images have been acquired on 2015 February 18 in five filters: Near-UV, Blue, Orange, Red, Infrared. The comet was at about 2.3 AU from the Sun and the phase angle was about 85 deg. The spacecraft was at 220 km from the comet and the NAC's field of view was about 9 km around the nucleus.

We constructed a map of radial distance from the nucleus limb and computed the average coma surface brightness over an azimuthal region generally free of gas jets and ghosts. Assuming that in the broad NAC filters (mostly ~50 nm) the contribution of the gas emission is small with respect to the dust contribution, we derived the colors of the dust coma. We found that the coma has an average reddening of 19% per 100 nm in the wavelength range 360 – 649 nm, the closest to the 375 – 610 nm range obtained with WAC filters.

However, NAC imaging is not available concurrent with most WAC gas observations, and it may or may not sample the relevant region of the coma given the instrument's small field of view. Therefore, we devised a second, empirical method to determine the reddening of the dust in the coma. We assumed that $NH_2$ emission comes predominantly from fragment species in the coma. Thus, for the images of 67P acquired with the $NH_2$ filter the underlying continuum subtraction was done by varying the continuum removal factor until jets disappeared from the $NH_2$ filter image leaving a rather isotropic $NH_2$ coma. This method typically yields factors that correspond to 17–20% per 100 nm between 375 and 610 nm, somewhat larger than reported from ground-based observations of the comet in previous apparitions (11 ± 2 % per 100 nm between 436 and 797 nm; Tubiana et al. 2011) but consistent with ground-based observations acquired during this apparition (20% between B and V; Snodgrass et al. 2015). We therefore used a constant reddening of 18% per 100 nm for the data discussed in this paper.

*3.2 Column Densities and Production Rates*

We identified observations in each sequence that were acquired at the same diurnal phase, extracted surface brightness profiles in two directions, and measured the surface brightness at a fixed position 3 km in the sunward direction above the surface.

When the formation and excitation processes of the fragment species are known, the measured surface brightnesses *S(x,y)* can be converted into column densities $N_{col}$ by assuming fluorescence efficiencies $g$ of each species:

$$N_{col}(x,y) = \frac{4\pi \cdot S(x,y) \cdot \Delta\lambda}{E_p \cdot g} \quad [4]$$

where $\Delta\lambda$ is the FWHM of the filter (Table 2) and $E_p$ the energy of the photons at the wavelength of the emission feature considered. [OI] surface brightnesses can be converted into $H_2O$ column densities using reaction rates for prompt excitation of the $^1S$ and $^1D$ states (Bhardwaj & Raghuram 2012), weighed by the branching ratios of the transitions leading to emission at 630 nm (the 2nd line of the doublet at 636 nm is at the edge of the OI filter where the transmission is below 6% of the peak transmission; see Fig. 1b). The CN, OH, and NH surface brightnesses were converted into column densities using published fluorescence efficiencies (Table 4). We have calculated production rates using a standard Haser model for easier comparison with other observations. Assumed outflow velocities and lifetimes for parents and daughters can be found in the Appendix (Table A1). To better evaluate the column densities, we also calculated production rates with a modified Haser model that takes the gas acceleration, collisional quenching of the long-lived 1D state of the oxygen atom, and the effect of the oxygen atom moving out of the field of view before it can decay to the ground state into account. This model is described in Appendix A.

*3.3 Uncertainties*

The results are subject to several possible systematic uncertainties. The absolute calibration of OSIRIS is better than 1% for most filters, but the calibration constant for the OH filter (and continuum filters <300 nm) has an uncertainty of ~10% (Tubiana et al. 2015). Bias levels are temperature dependent and have gradually changed over time because the spacecraft approached the Sun. From the 16Cyg observations (where we can see the background), we estimate that the bias level is now constrained to within 1 DN. Because the bias is individually determined for each hardware configuration, the error remains 1 DN independent of binning. The resulting gas detections typically had a SNR of 4 or better per 4x4 pixel at 100 pixels from the nucleus. The continuum removal is the largest systematic and statistical uncertainty in our data analysis. For example, in the data acquired in March, at a distance of 100 pixels, i.e., 0.82 km, from the nucleus, the continuum contributes 10% to the total signal in CN, ~20% in OI, ~30% in OH, and as much as 65% of the signal in the $NH_2$ filter. We have also tried to optimize the quality of the continuum removal factors by averaging repeated observations of 16Cyg, and these coefficients are now constrained within 5 – 10% (Table 3). As discussed above, we only use the UV375 filter to observe the continuum emission, and have to infer the color of the coma. We do not account for spatial gradients and temporal variations in the color of the coma. Assuming that the reddening is typically between 0 – 30%, differences in reddening could result in uncertainties of <1% (CN), 8% (OI), 5% (OH), and 30% ($NH_2$) in the resulting pure gas emission after continuum subtraction. .

Uncertainties in the assumed fluorescence efficiencies can affect the results systematically. Even more so, the use of those factors rests on the premise that the dominant processes are photo-processes, which we deem not be the case (Sec. 5).

## 4. RESULTS

Continuum and contaminant removed narrowband images for March and June are shown in Fig. 2, along with contextual images acquired with the 375 nm filter. All images are oriented such that the direction to the Sun is upward. In each frame, the entire field of view of the WAC is shown, which sides corresponds to 16 and 46 km projected at the nucleus, respectively. In each image a circular feature can be seen on the right side of the nucleus (saturated pixels are masked black in our image processing). These 'ghosts' are caused by internal reflections from optical elements and are always placed on the detector at the same place with respect to the nucleus. A smaller, weaker artifact can be seen in the same direction directly next to the nucleus. Three remnant pinholes can clearly be seen in the OH images.

All continuum images show multiple bright, collimated jets on the sunward side of the nucleus. In the top row, the March data, those jets are not present in any of the continuum-subtracted gas images. At all epochs we see a plume-like morphology perpendicular to the sunward direction. This feature is seen in the OI and CN filter, but not in the OH, NH, and $NH_2$ filters, where the observed morphology is less pronounced and enhanced towards the Sun (Fig. 2).

The continuum subtracted $NH_2$ images have poor SNR and show a broad distribution with little structure. For the data acquired in June and July, while assuming the same reddening of 18% per 100 nm (Sec. 3.1.2), the jets are over-subtracted in the OI data and cannot be entirely removed from most of the other filters. We believe this to be a consequence of our approach to the continuum subtraction (i.e. assuming a constant color throughout the inner coma). For each of the images in Fig. 2, the color scale of the look up table was adjusted to best emphasize the morphology. It is clear that while the morphology in the OH and NH images remains isotropic and diffuse, the cone in the [OI] and CN images becomes much fainter and less defined over time.

Radial surface brightness profiles were extracted from a 21-pixel wide box between the left limb of the nucleus and the edge of the frame. We extracted surface brightnesses in both the horizontal and vertical (sunward) direction, starting at the nucleus (orientation as in Fig. 2). The 1-sigma uncertainty was estimated from the standard deviation within the 21-pixel wide box. The resulting continuum subtracted profiles are shown in Fig. 3. The first two epochs show the best SNR. The surface brightness decreased by a factor of 5-10 after February, resulting in visible poorer SNR. For the observations in April, May, and June, data within 1 km from the surface seems unreliable, with the exception of CN. $NH_2$ is consistently very noisy (as the plots have a logarithmic scale, negative values are not shown), and it was probably not detected in the June and July 2015. On all six epochs, there is a clear difference between the shape of the profiles of OH and NH, which show a 1/r drop-off with distance, and the [OI] and CN profiles, which are flat within the first ~8 km from the surface. In January and March, the OH profiles in the sunward direction are very similar to those in the perpendicular direction. After that, the emission on the sunward side of the coma is much stronger (factor 2) than that in the direction orthogonal to the comet-sun line. This asymmetry is also present in the NH profiles, albeit less pronounced. In the OI and CN profiles, the emission is somewhat stronger in the orthogonal direction because the plumes visible in Fig. 2 extend to both extraction directions. On the two epochs where we have a good NH2 detection, January and March, its profiles are very flat and very different from the NH profile.

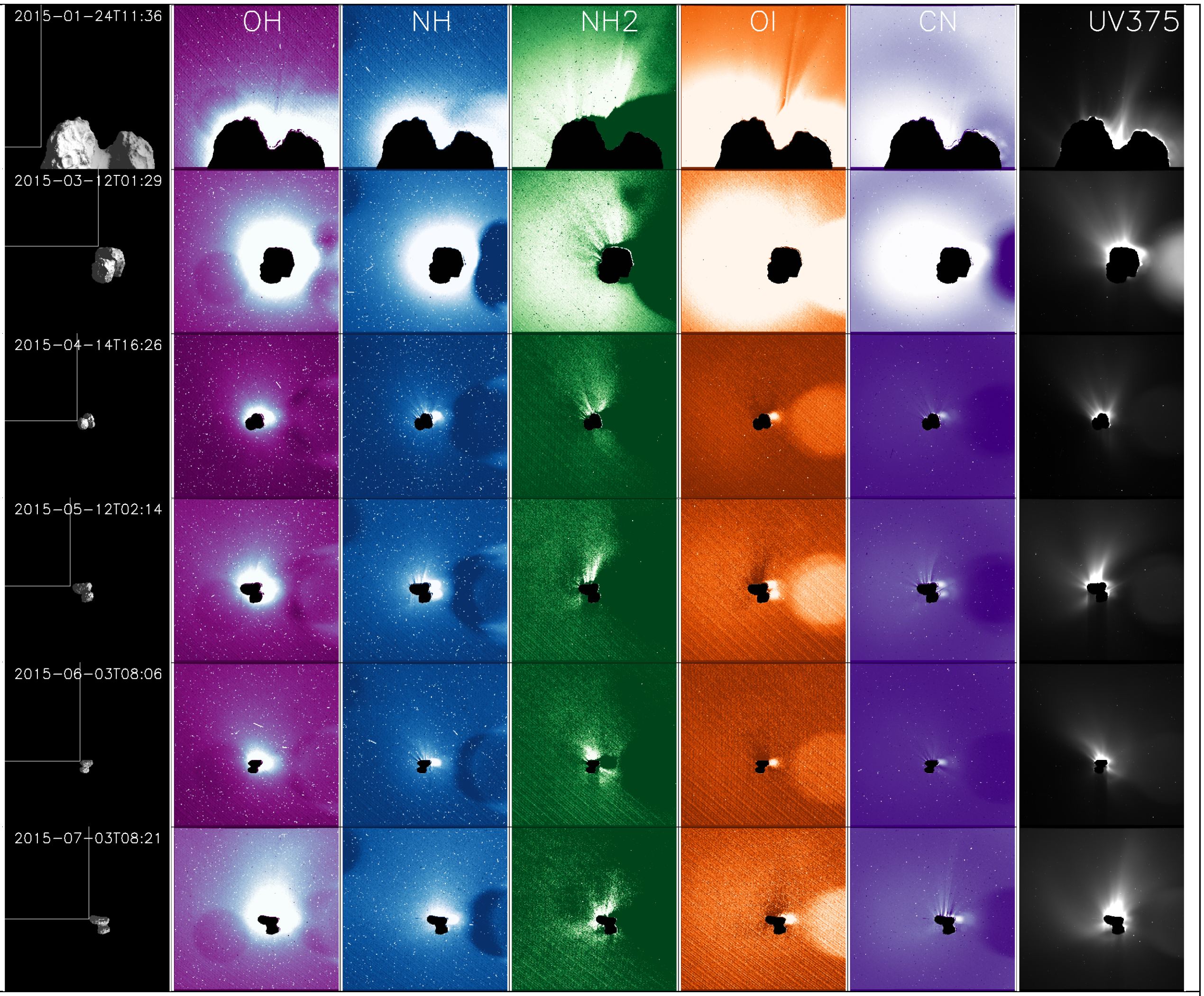

**FIG 2.** Left to right: shape model image for reference, continuum subtracted images acquired with the OH, NH, $NH_2$, OI, and CN filters, and a long-exposure UV375 filter image for comparison. In the orientation used here, the Sun is always toward the top, and the main reflection ghost can be seen to the right of the nucleus. All images are the entire field of view of the WAC (11.4 x 12.1 degrees). The color scale is different for every filter, but is kept constant throughout all epochs within each filter.

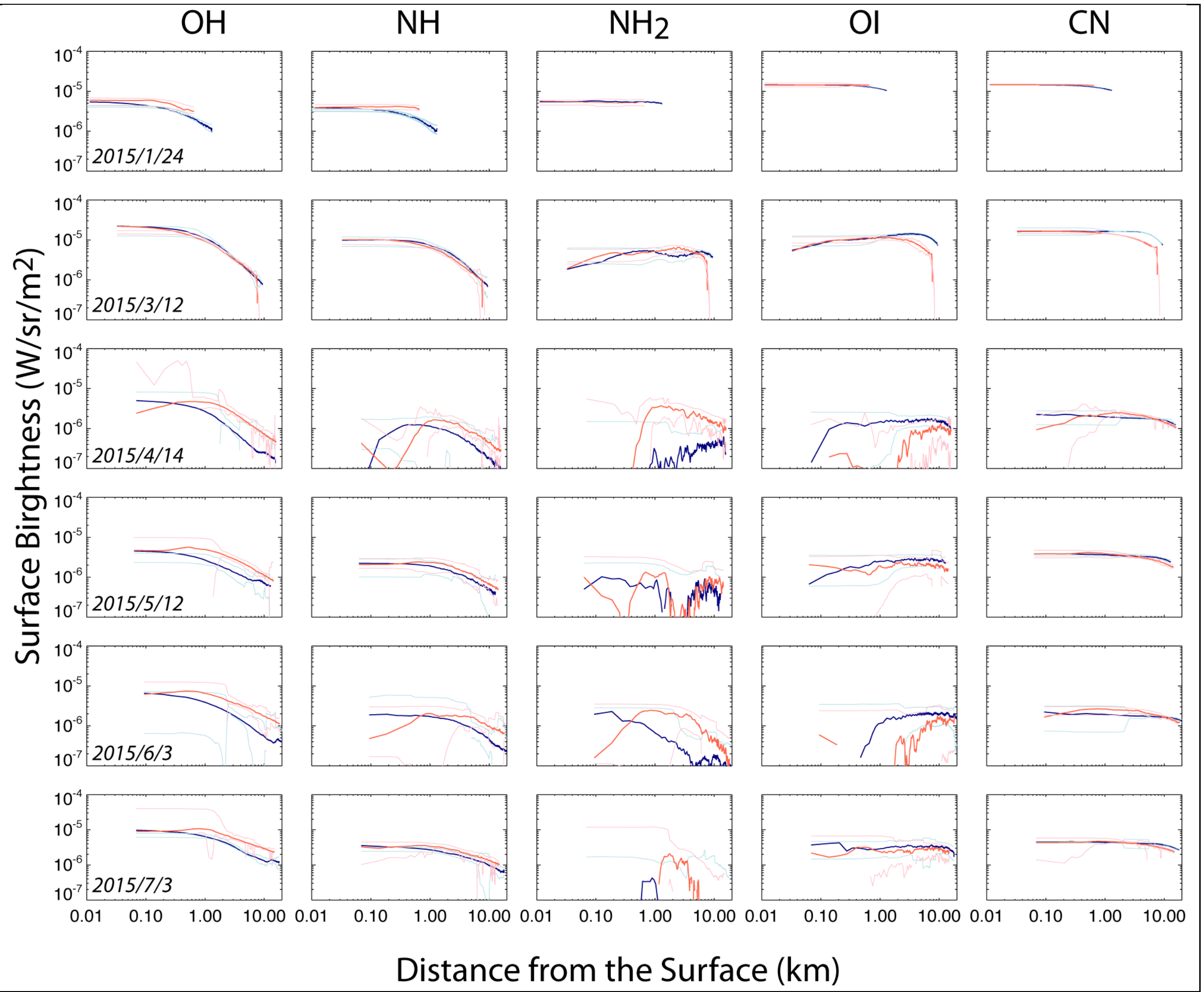

**Fig. 3** – Surface brightness profiles. Blue lines are profiles in the horizontal direction ('plume-ward'); red lines are profiles in the vertical direction (sunward; cf. Fig. 2A). The thin, lighter shade lines (pink and cyan) indicate the 1-sigma uncertainties.

## 5. EMISSION PROCESSES

There are numerous inconsistencies in our results if we assume standard cometary physics. First, column densities and production rates derived from OSIRIS images are much higher than those measured by other instruments on board *Rosetta* (MIRO, VIRTIS, and ROSINA/COPS; Bieler et al. 2015a; Bockelée-Morvan, D. et al., 2015; Fougere et al., 2016;). Assuming photodissociation of $H_2O$ as the main source of formation and prompt emission by atomic oxygen, and photodissociation and subsequent fluorescent excitation of OH, CN, and NH emission, we derived column densities and calculated global production rates using standard Haser model (Fig. 4, Tables 5 and 6).

There are currently no contemporaneous measurements available of the abundance of CN and NH. Abundances of $NH_3$ and HCN were 0.06% and 0.09% with respect to water, measured by *Rosetta's* ROSINA instrument at 3.1 AU on the sunward side of the comet (Le Roy et al. 2015). Fink (2009) measured $NH_2$ and CN abundances of 0.19% and 0.15% from

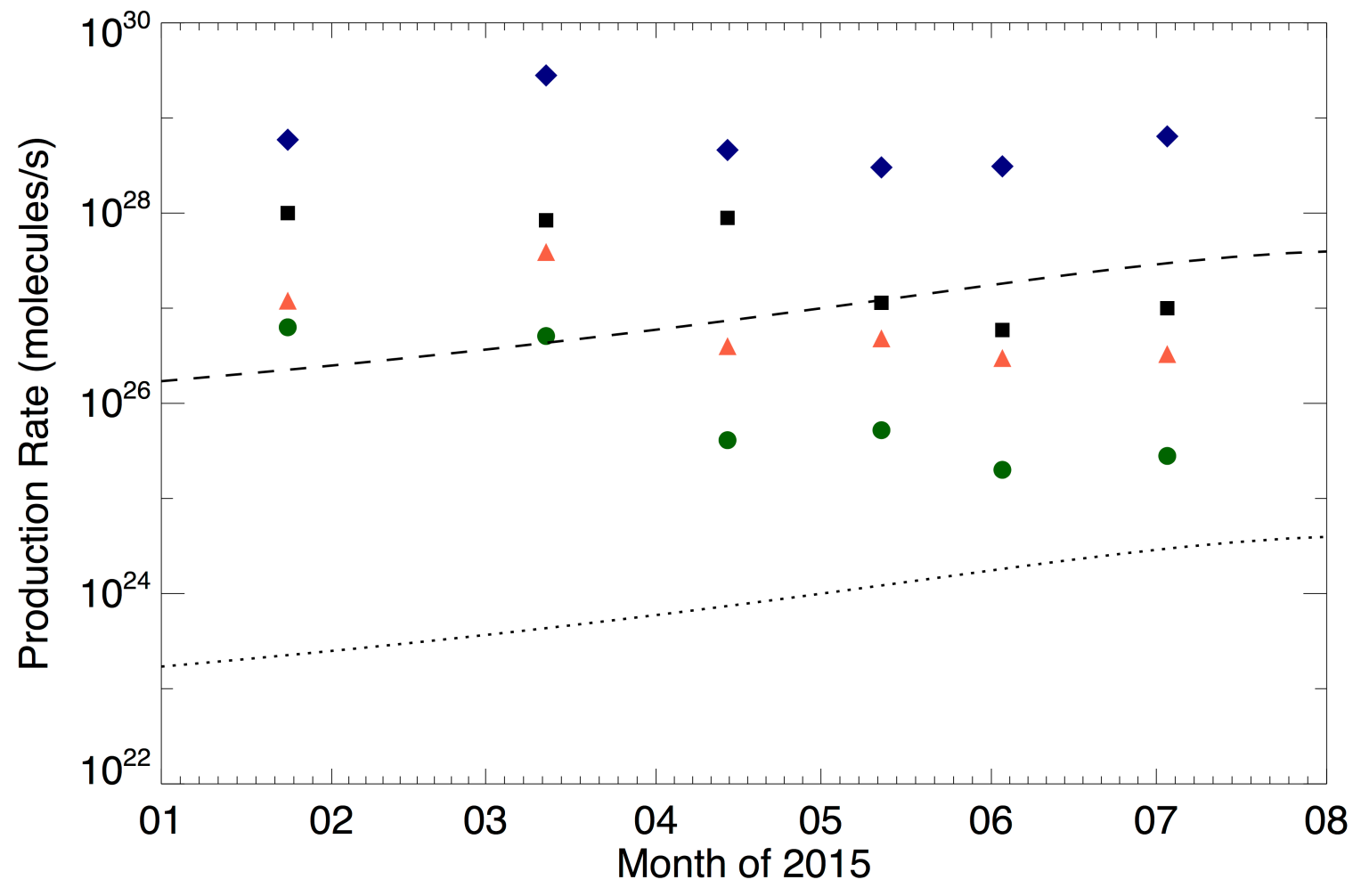

**FIG. 4A** – Water production rates derived from OH (blue diamonds) and [OI] (black squares), CN production rates (green circles), and NH production rates (red triangles), based on the assumption that photo dissociation and fluorescence are the driving destruction and excitation mechanism. The dashed line indicates $H_2O$ production rates derived from MIRO measurements (Fougere et al., 2016); the dotted line indicates expected production rates of $NH_3$ and HCN assuming fixed abundances of 0.1%.

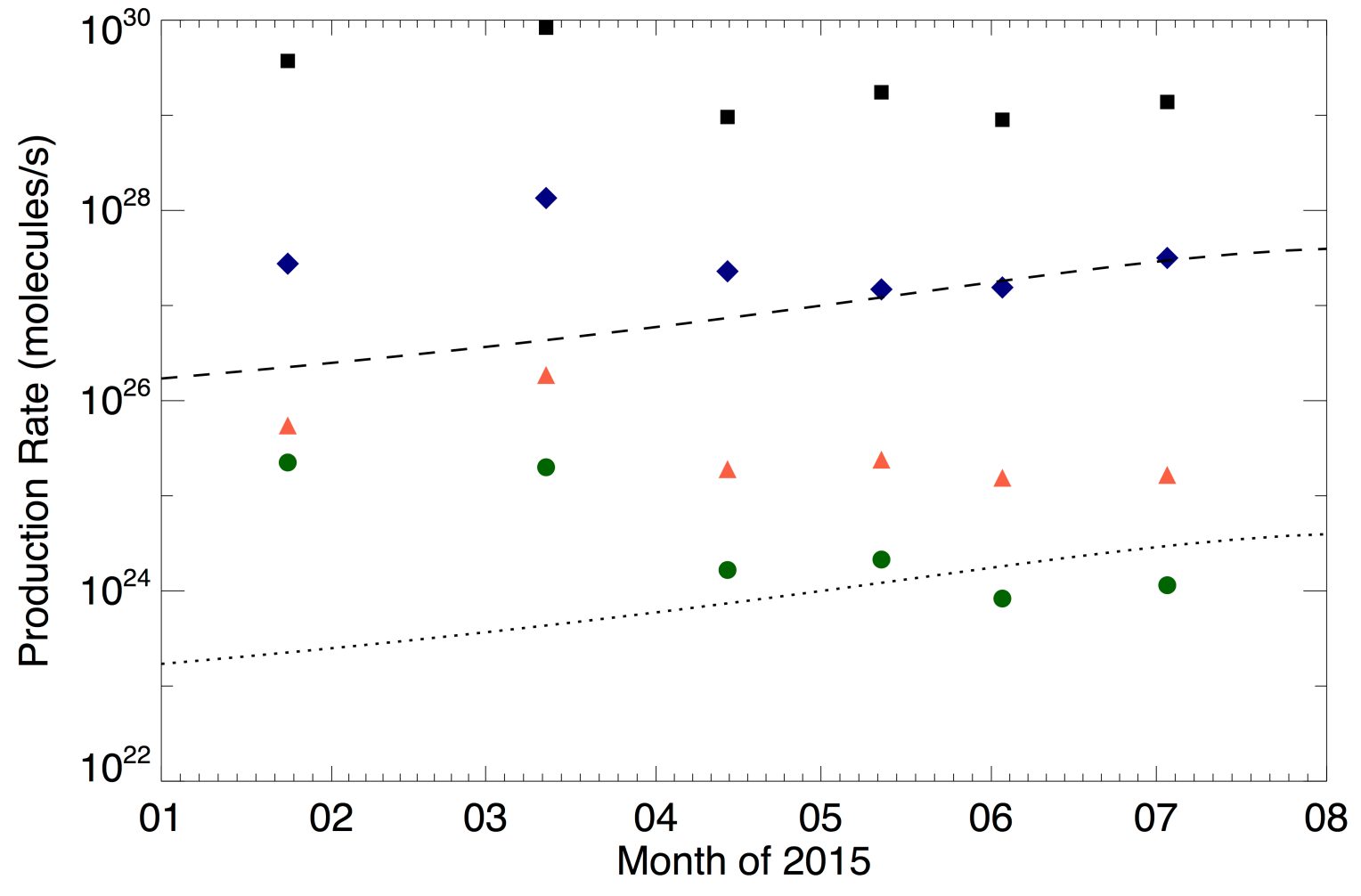

**FIG. 4B** – As Fig 4A but production rates are now calculated using an enhanced model that includes acceleration in the inner coma. $H_2O$ production rates derived from [OI] emission include quenching and transport effects.

the ground at 1.35 AU from the Sun post-perihelion during the 1995 apparition. To calculate the expected surface brightness levels, we assumed abundances of 0.1% for HCN and $NH_3$, and water production rates from Fougere et al. (2016). As shown in Fig. 4a, water production rates derived from the OH observations using the standard Haser model are initially more then a factor 300 larger then expected. Production rates derived from CN and

**Table 4 –** Photon production efficiencies based on photodissociation rates (OI) and fluorescence models (OH, CN, NH) in units of photons/molecule.

| DATE | OI[1] | OH[2] | CN[3] | NH[4] |
|---|---|---|---|---|
| | (photons/molecule/s) | | | |
| **1/24/2015** | 9.95E-8 | 7.13E-5 | 0.010 | 1.89E-3 |
| **3/12/2015** | 1.35E-7 | 9.21E-5 | 0.014 | 2.56E-3 |
| **4/14/2015** | 1.72E-7 | 1.18E-4 | 0.018 | 3.29E-3 |
| **5/12/2015** | 2.20E-7 | 1.61E-4 | 0.023 | 4.20E-3 |
| **6/3/2015** | 2.85E-7 | 1.95E-4 | 0.029 | 5.09E-3 |
| **7/3/2015** | 3.59E-7 | 1.43E-4 | 0.041 | 6.68E-3 |

**References:** 1) Bhardwaj & Raghuram 2012; 2) Schleicher 2010; 3) Schleicher & A'Hearn 1988; Kawakita & Watanabe 2002; 4) Kim et al. 1989.

NH surface brightnesses are also large by factors of 1000. This changes dramatically when we assume a more realistic velocity (eq. A2) in the inner coma, see Fig. 4b. Derived production rates for $H_2O$ derived from OH are now about one order of magnitude too large, those of NH and CN parents by about two orders of magnitude. For OH and CN, the July results are consistent with expected production rates. NH emission remains about a factor of 10 too high.

Second, with the WAC we expect to observe two water photolysis products, OH and OI. Using the standard Haser model, water production rates derived from OH seem consistently larger than those derived from [OI], by a factor of 6 in January and by a factor of 30 in July. This situation is not resolved by our appended coma model because both there is no change in the relative column densities of OH and OI if both are assumed to come from $H_2O$.

Third, all production rates drop significantly between March and June, 2015. This variation is much larger than the diurnal variation of the total water production (~25%; Gulkis et al. 2015) and not consistent with the observed trend of increasing production rates with decreasing heliocentric distance.

Fourth, the luminosity profiles are at odds with the observed morphology. Surface brightness profiles of parent species usually decrease with ~1/r close to the nucleus, whereas those of fragment species have shallower slopes (c.f. Combi et al. 2004). The OSIRIS surface brightness profiles (Fig. 3) however are flat for [OI] and CN, for which the morphology suggests a prompt excitation process. In contrast, the morphology seen in the OH and NH filters – a symmetric distribution around the nucleus – is typical for a fragment species that gets a significant vectorial kick upon photodissociation of a parent species. The morphology of emission in the OI and CN images resembles the projection of a cone of gas and indicates a parent dissociation process that produces these fragments directly into an excited state.

We therefore conclude that by adjusting our models to better describe the physical processes in the inner coma we can explain some of the observations, but that the differences between our observations and model results indicate that photodissociation and fluorescence are not the dominant processes resulting in the OH, [OI], CN, and NH emission observed in the inner coma. Instead, the fragments might be fragments from different parent species and/or formed by other processes. We will discuss this in further detail below.

**Table 5 –** Surface brightness measured at 1 km above the surface in the horizontal direction (orthogonal to sunward direction). Column densities and production rates are derived assuming photo processes and are only given to show the discrepancy between observations and expected production rates (Sec. 5).

| | **Surface Brightness ($10^{-6}$ W/m$^2$/sr)** | | | | | **Log column densities (molec./m$^2$)** | | | |
|---|---|---|---|---|---|---|---|---|---|
| **Date** | **OI** | **OH** | **CN** | **NH** | **$NH_2$** | **OI** | **OH** | **CN** | **NH** |
| **1/24/2015** | 11.5±0.2 | 1.5±0.2 | 11.4±0.2 | 1.4±0.2 | 5.3±0.3 | 21.7 | 17.6 | 16.5 | 16.2 |
| **3/12/2015** | 12.1±1.0 | 11.0±1.3 | 16.5±0.2 | 7.4±0.5 | 4.8±1.4 | 21.6 | 18.4 | 16.5 | 16.8 |
| **4/14/2015** | 1.3±0.8 | 2.7±0.2 | 2.0±0.1 | 1.1±0.8 | 0±1 | 20.5 | 17.6 | 15.4 | 15.9 |
| **5/12/2015** | 2.4±0.9 | 2.7±0.4 | 3.7±0.2 | 2.0±0.5 | 0.9±1 | 20.6 | 17.5 | 15.6 | 16.0 |
| **6/03/2015** | 1.6±1.0 | 3.8±0.2 | 2.0±0.3 | 1.7±0.5 | 0.8±0.8 | 20.1 | 17.6 | 15.2 | 15.8 |
| **7/03/2015** | 3.1±1.6 | 6.4±0.3 | 4.4±0.3 | 2.7±0.3 | 1.1±2 | 20.5 | 17.9 | 15.4 | 15.9 |

**Table 6 –** Production rates derived assuming photo processes and are only given to show the discrepancy between observations and expected production rates (Sec. 5). Results are shown for a standard Haser model, for an enhanced model that includes acceleration (incl. OI[a]), and for a model that includes quenching and transport of OI (OI[b]) The production rates are all for the assumed parents of the observed fragments, the label indicates from what fragment they were derived (i.e. OI for $H_2O$, OH for $H_2O$, CN for HCN, and NH for $NH_3$).

| | **Standard Haser** | | | | **Enhanced model** | | | | |
|---|---|---|---|---|---|---|---|---|---|
| | **Log prod. rates (molec./s)** | | | | **Log prod. rates (molec./s)** | | | | |
| **Date** | **OI** | **OH** | **CN** | **NH** | **OI[a]** | **OI[b]** | **OH** | **CN** | **NH** |
| **1/24/2015** | 28.0 | 28.8 | 26.8 | 27.1 | 27.7 | 29.6 | 27.4 | 25.4 | 25.7 |
| **3/12/2015** | 27.9 | 29.5 | 26.7 | 27.6 | 27.6 | 29.9 | 28.1 | 25.3 | 26.3 |
| **4/14/2015** | 27.9 | 28.7 | 25.6 | 26.6 | 26.6 | 29.0 | 27.4 | 24.2 | 25.3 |
| **5/12/2015** | 27.1 | 28.5 | 25.7 | 26.4 | 26.7 | 29.2 | 27.2 | 24.3 | 25.4 |
| **6/03/2015** | 26.8 | 28.5 | 25.3 | 26.5 | 26.5 | 29.0 | 27.2 | 23.9 | 25.2 |
| **7/03/2015** | 27.0 | 28.8 | 25.5 | 26.5 | 26.7 | 29.1 | 27.5 | 24.1 | 25.2 |

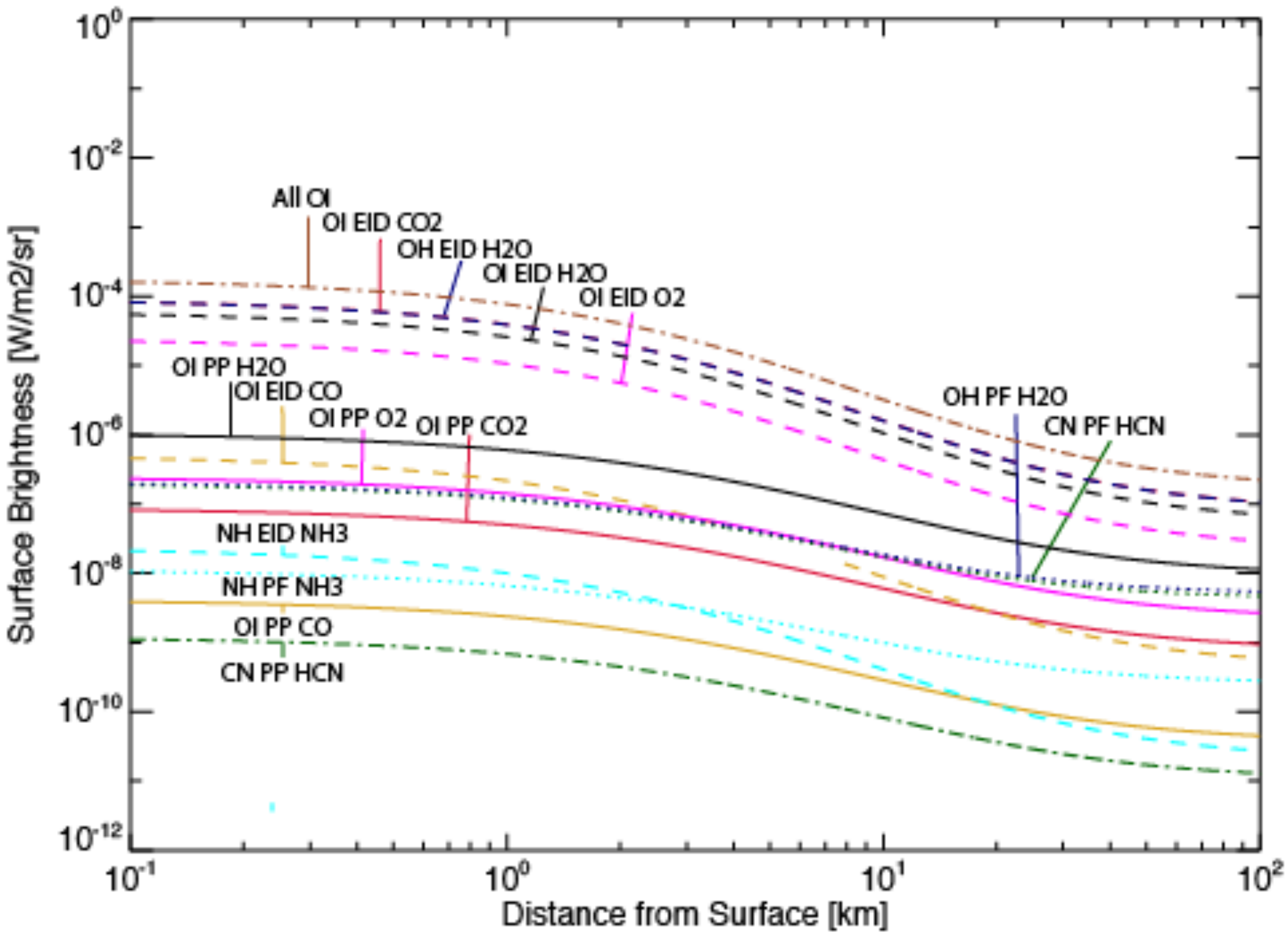

**FIG. 5A** – Modeled surface brightnesses from different processes in the coma for the Jan 24, 2015 observations. EID –electron impact dissociation (dashed lines). PP – Photodissociation, Prompt excitation (solid lines). PF – Photodissociation and subsequent fluorescent emission (dotted lines). Colors indicate the parent species: Black and blue – $H_2O$ products; Red – $CO_2$ products; Orange – CO products; Magenta – $O_2$; Green – HCN. Cyan – $NH_3$. The dash-dotted brown line shows the sum of all [OI] emission.

### *5.1 Water fragments: OH and OI*

As concluded above, the emission from [OI] and OH cannot be explained by photodissociation of $H_2O$, followed by prompt emission of [OI] or by fluorescence of OH. The surface brightness profile of OH suggests its emission might be the product of a process that produces OH directly in the $A^2\Sigma^+$ state. However, the difference between the [OI] and OH morphology indicates that at least part of the emission of the two fragments is not related.

From the morphology of the [OI] emission we concluded that it is the product of a process that directly produces atomic oxygen in an excited state. Like $H_2O$, photodissociation of $CO_2$ and CO produces OI in the $^1D$ and $^1S$ states, resulting in [OI] emission at 630 nm. Abundances of $CO_2$ and CO vary greatly between the summer and winter hemispheres. Here we assume that the sunlit side dominates the total gas production rate; there, abundance ratios from the 'summer hemisphere' apply, which are $H_2O$: $CO_2$ : CO = 100 : 2.7 : 2.5 (Le Roy et al. 2015). Assuming $CO_2$ and CO formation rates into OI $^1D$ and $^1S$ from (Bhardwaj & Raghuram 2012), photodissociation of $CO_2$ and CO contributes 10% and <1% to the [OI] 630 nm emission compared to $H_2O$ (100%).

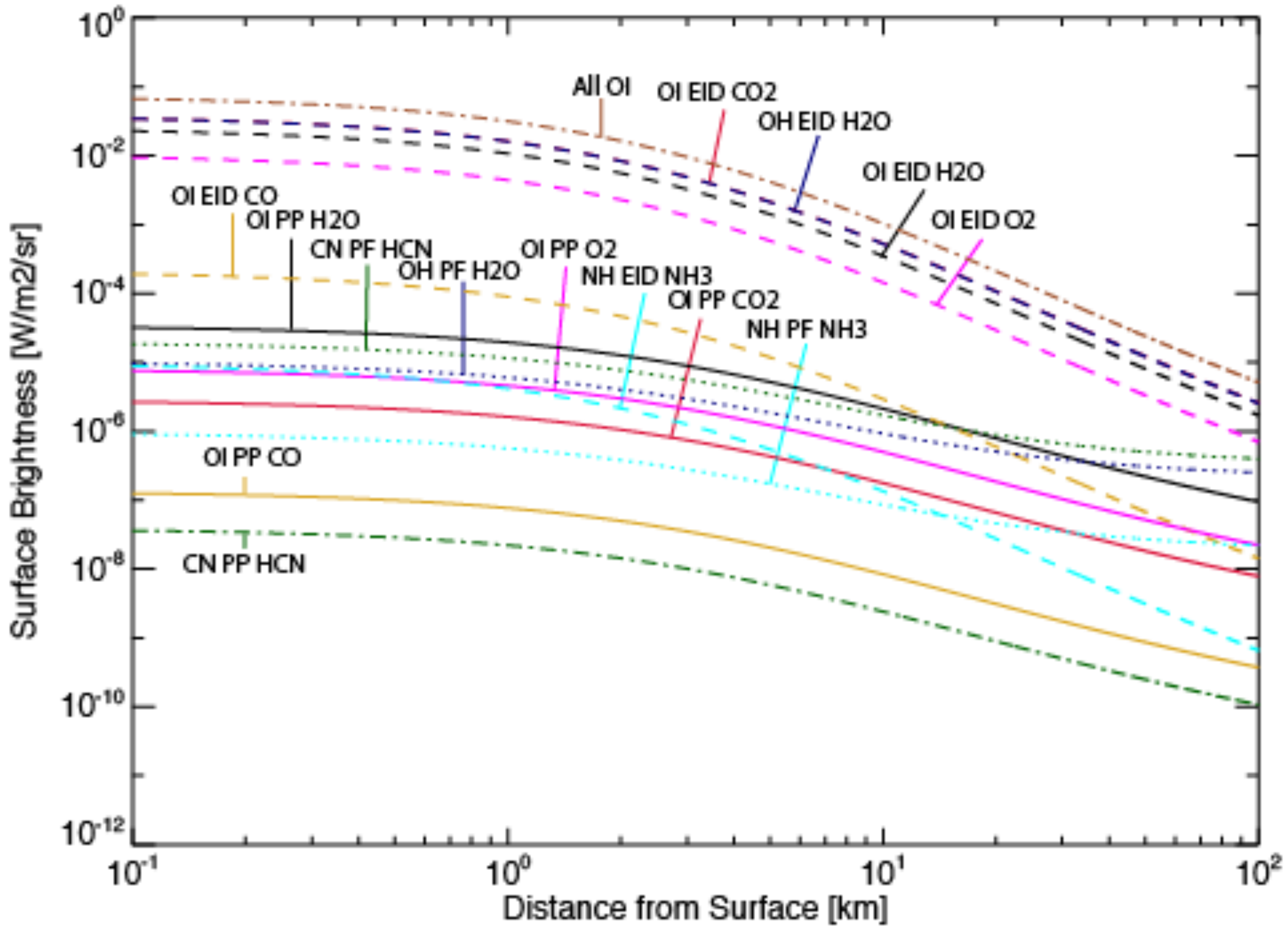

**FIG. 5B** – As Fig. 5A above but for the July 3, 2015 observations.

The ROSINA instrument reported unexpectedly high abundances of $O_2$, with an average of $O_2/H_2O$ = 3.7 ± 1.5 %, with local abundances as high as 10% (Bieler et al. 2015b). The distribution of $O_2$ in the coma suggested it is released by the nucleus and that its release is correlated to the outflow of water. Because the photodissociation of $O_2$ into OI $^1D$ is very efficient and at these abundances (Huebner et al. 1992), it can contribute as much as an additional 25 – 60% to the [OI] 630 nm emission from $H_2O$. Thus while the photodissociation of $CO_2$, CO, and $O_2$ molecules combined may produce as much [OI] emission as the photodissociation of $H_2O$, it cannot explain the factors of 20 – 40 of flux excess observed between January and March, nor can it explain the OH observations.

We then considered several processes that might produce the high observed surface brightness of both OH and [OI] directly from $H_2O$, including dissociative recombination of $H_2O^+$, electron excitation of OI, and sputtering of water ice (c.f. Bhardwaj & Raghuram 2012) but none of those have reaction rates that exceed that of photo processes. Feldman et al. (2015) concluded that electron impact dissociation of $H_2O$ vapor produced HI and OI emission observed by *Rosetta*/Alice in the Far-UV. Electron impact dissociation produces OI in the $^1D$ and $^1S$ states, and OH in the $A^2\Sigma^+$ state. Those reactions typically have appearance thresholds between 10-20 eV, suggesting that they are driven by the large population of suprathermal electrons observed in by *Rosetta*'s Ion and Electron Sensor (IES; Clark et al. 2015). While electrons with energies of 100 eV and larger were observed, the distribution falls off steeply >30 eV. At these impact energies, the cross section for electron impact production of OH ($A^2\Sigma^+$) is 7 x $10^{-22}$ $m^2$ (Avakyan & al 1998) and that of OI ($^1S+^1D$) is 6 x $10^{-23}$ $m^2$ (Bhardwaj & Raghuram 2012).

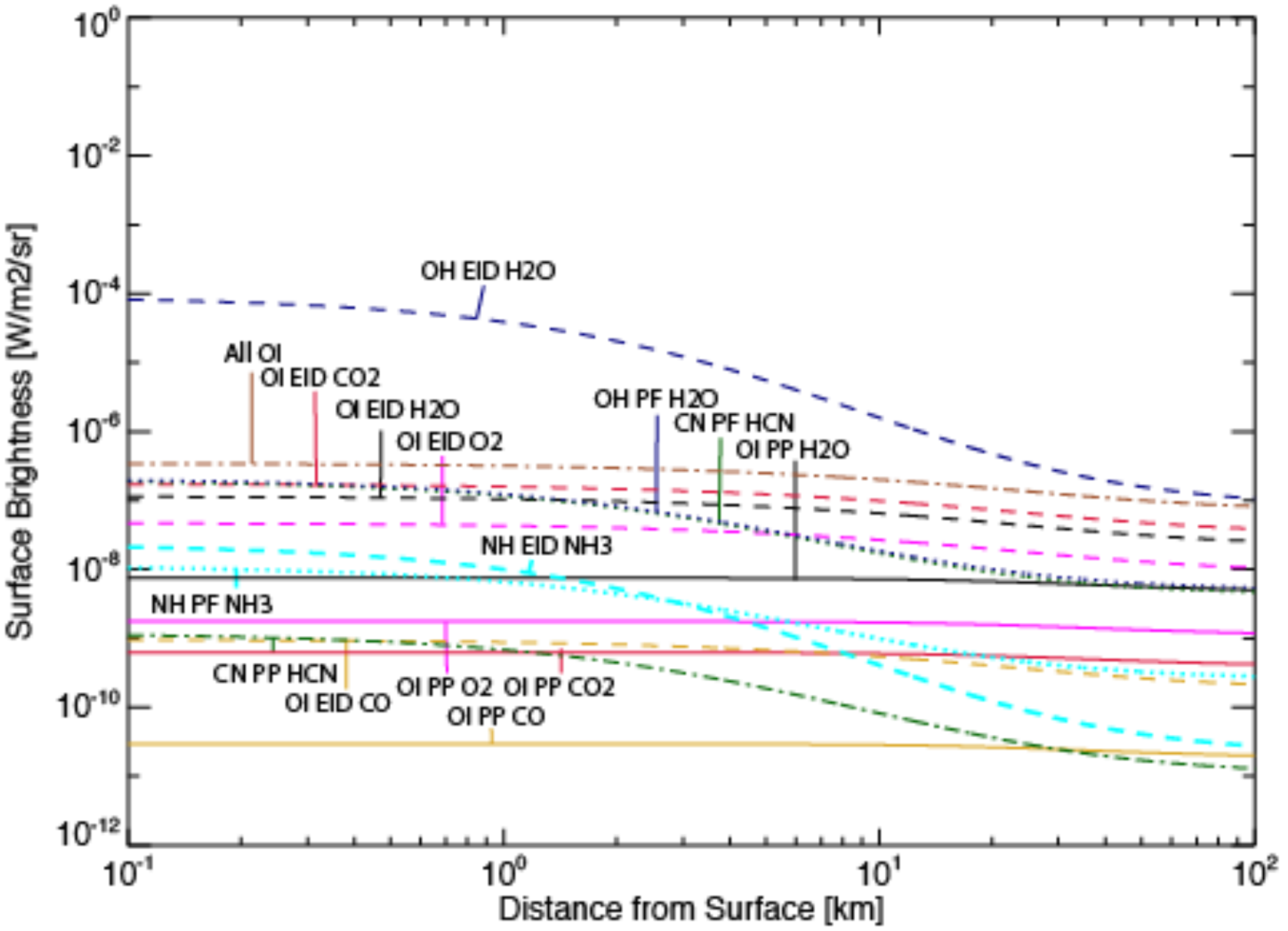

**FIG. 5C** – Modeled surface brightnesses from different processes in the coma for the Jan 24, 2015 observations, but including the effect of collisional quenching and transport. EID – electron impact dissociation (dashed lines). PP – Photodissociation, Prompt excitation (solid lines). PF – Photodissociation and subsequent fluorescent emission (dotted lines). Colors indicate the parent species: Black and blue – $H_2O$ products; Red – $CO_2$ products; Orange – CO products; Magenta – $O_2$; Green – HCN; Cyan – $NH_3$. The dash-dotted brown line shows the sum of all [OI] emission.

To test whether electron impact dissociation can explain the observed surface brightness we added electron impact processes to our model (Appendix A). In brief, we assumed a spherically outgassing nucleus with water production rates from the empirical trend reported by Fougere et al. (2016), and relative abundances of 3% for CO and $CO_2$, 0.1% for HCN and $NH_3$. For the electrons, we assumed a radial distribution that decreased with the inverse of the distance to the nucleus based on measurements by the *Rosetta*

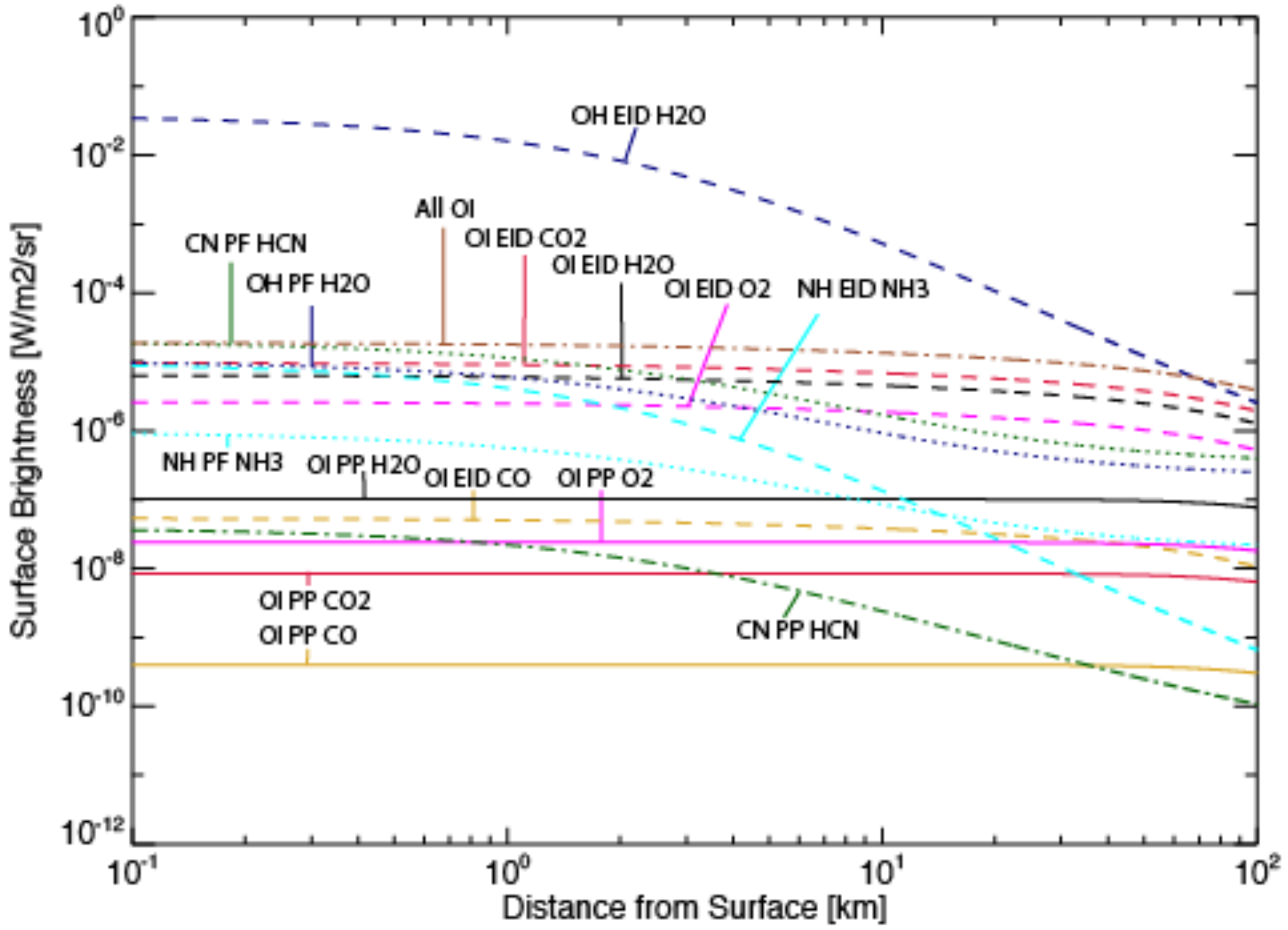

**FIG. 5D** – As Fig. 5C above but for the July 3, 2015 observations.

Plasma Consortium's (RPC) Langmuir- and Mutual Impedance Probes. For the electron density, we used measurements with the Electron and Ion Sensor (IES) (Broiles et al. in press) which we assumed to increase linearly with the production rate and quadratically with the decrease in heliocentric distance. In addition, we included the effects of quenching of the OI $^1D$ state by collisions with $H_2O$ molecules as well as the effect of transport as the OI atom moves out of the field of view before it can decay to the ground state.

The results of the model are shown in Fig. 5. We find that at 1 km from the surface, electron impact dissociation of $H_2O$ can produce up to a factor of 10 more [OI] 630 nm emission than the photodissociation of $H_2O$ into excited OI $^1D$, and at least two orders of magnitude more OH emission than the photodissociation of $H_2O$ and subsequent fluorescent excitation of OH by sun light. However, the morphology and surface brightness profiles seen in the OI and OH filters is very different, which is surprising when we assumed they are both the product of electron impact dissociation of $H_2O$. In the OI images there is a clear plume visible to the left, which is entirely absent in OH images. If the main emission in the plume originated from electron impact dissociation of $H_2O$, it should also be present in OH emission. If electron impact dissociation drives the emission of [OI] and OH in the coma, then the atomic oxygen emission in the plume is probably produced from a molecule other than $H_2O$.

We are not aware of experimental cross sections for the production of OI ($^1$D), but based on theoretical cross sections, Bhardwaj & Raghuram (2012) suggest that reaction rates for producing OI ($^1$D) from $CO_2$ are as much as 40 times larger than reaction rates producing OI ($^1$D) from $H_2O$. The reaction rates for the production of excited oxygen atoms from CO are much lower than those of $H_2O$. Assuming these electron impact dissociation cross sections, a $CO_2/H_2O$ abundance of 3% would imply that 60% of the observed [OI] emission comes from $CO_2$. This large contribution by electron impact dissociation of $CO_2$ explains the differences between both the morphology and surface brightness profiles of the [OI] and OH emission. The plume is visible in [OI] but not in OH. Its [OI] surface brightness at the core is ~2x that at similar distances in the ambient coma, suggesting a local enhancement of the $CO_2$ abundance.

We do not expect our model to provide a fully realistic description of the inner coma. Predicted OH surface brightnesses are higher than observed, and quenching and transport may remove as much as factor 100 from the detectable [OI] surface brightness (Fig. 5a and Fig. 5c). In particular, the asymmetric outgassing of 67P will affect the impact of these processes. However, our results do confirm that electron impact dissociation can indeed explain the emission observed in OSIRIS' OI and OH filters from January through April 2015.

### *5.2 Ammonia fragments: NH and $NH_2$*

The photodissociation of $NH_3$ results in the production of $NH_2$ (96%) and NH (~3%); most of the NH is thus normally a granddaughter product of $NH_3$ (Huebner et al. 1992). Electron impact dissociation of $NH_3$ has been relatively well studied and cross sections for the production of $NH_2$ ($\tilde{A}\ ^2A_1$) and NH ($A^3\Pi_1$) at 100 eV are both ~2-3 x $10^{-22}$ $m^2$ (Müller & Schulz 1992), i.e. three times smaller than the production of excited hydroxyl (Table A1). To our knowledge, a detailed study of the excitation of $NH_2$ through this process is not available, leaving the relative intensities of the $NH_2$ bands that contaminate several filters an open question (Sec. 3.1.1). Surprisingly, the surface brightness profiles of $NH_2$ and NH are very different (Fig. 3), which is inconsistent if the emission from both fragments is produced by electron impact dissociation of $NH_3$. We used our model to evaluate the surface brightness expected from different emission processes. In all observations, the profile of NH resembles that of OH, suggesting that the same physical process causes the emission. The profile of $NH_2$ resembles that of OI and CN in January and March, and was probably not detected after that.

Assuming an abundance of 0.1% with respect to $H_2O$, we included electron impact excitation and fluorescent emission in our model. Photodissociation of NH3 into NH2 and NH cannot explain the emission for any of the observations (Fig. 4b), nor does electron impact dissociation (Fig. 5c), which both fall about a factor 100 short. Upon further investigation, the only plausible explanation for the NH filter observations appears to be emission from the $OH^+$ ($A^3\Pi$ – $X^3\Sigma^-$ 0-0) band. Electron impact dissociation on $H_2O$ can indeed produce excited hydroxyl ions with a cross section of 8 x $10^{-24}$ $m^2$ at 100 eV (Müller et al.1993), or about 100 times smaller than that for the production of OH ($A^2\Sigma^+$). The emission cross sections for neutral and ionized OH both depend strongly on the electron impact energy and the ratio between the surface brightnesses in the NH and OH filters will therefore depends strongly on the electron temperature in the coma.

*5.3 CN*

The morphology of the CN emission resembles that of [OI] at all six epochs, and like the water group fragments, its surface brightness is higher than expected from photo processes, by as much as a factor of at least 200 (Fig. 4b). Prompt excitation of CN following photodissociation of HCN is not a very efficient emission mechanism (Fray et al. 2005; Bockelée-Morvan & Crovisier 1985), and our model indeed suggests that photodissociation of HCN and subsequent fluorescent excitation of CN would produce at least 50 times more light than prompt excitation of CN following the photodissociation of HCN. Following the discussion on the production of [OI] and OH emission, we then evaluated if electron impact dissociation of HCN into the CN ($B^2\Sigma$) state can drive the production of excited CN. Qualitative results confirm that this reaction channel is important in electron impact dissociation of HCN (Nishiyama et al. 1979), but cross sections are not available in the literature. Assuming cross sections and abundances similar to $NH_3$ reactions for HCN, we would expect surface brightnesses comparable to NH in January (Fig 5c), i.e. around $10^{-8}$ $W/m^2/sr$ at 1 km from the surface. Interestingly, owing to the large fluorescence efficiency of CN, emission levels from the photodissociation of HCN followed by fluorescent excitation of CN likely produces an order of magnitude more light than electron impact dissociation of HCN. Dissociative electron impact excitation of HCN cannot explain our observations.

This suggests that the light observed in the CN filter in January is produced by another species. According to Ajello et al. (1971) about 5% of the light emitted following ionizing electron impact excitation of $CO_2$ falls within the CN filter's passband. This also explains the similarity in morphology between the OI and CN filter images. At an impact energy of 30 eV, the emission cross section for the entire $CO_2^+$ ($\tilde{A}\,^2\Pi - X^3\Sigma^-$) band is 4 x $10^{-21}$ $m^2$, thus resulting in an emission cross section of 2 x $10^{-22}$ $m^2$, about 4x smaller than that of dissociative electron impact excitation of $H_2O$ into OH*, and about 10x smaller than the production of OI $^1D$ from the dissociative electron impact excitation of $CO_2$. We note that while the [OI] emission is strongly affected by quenching and transport effect but that the $CO_2^+$ emission is not, which may explain why [OI] surface brightnesses are comparable to those measured in the CN filter (Table 5). The expected $CO_2^+$ surface brightness can be estimated from the $CO_2$-to-[OI] profile in Fig. 5a and will be of order $10^{-5}$ $W/m^2/sr$, or about 100 times that of fluorescent emission from CN. Photodissociation of a parent molecule and subsequent fluorescent excitation of fragment CN can explain the observed surface brightness levels after May 2015, but the persistent plume morphology suggests that the observed emission may be a product of both CN fluorescence emission and $CO_2^+$ emission from electron impact.

## 6. TRENDS WITH PERIHELION DISTANCE

The production rates shown in Fig. 4 suggest three different epochs: increasing emission before March 2015 following the heliocentric trend observed by MIRO, then a sharp decrease between mid-March and June, 2015, followed by an increase after June 2015. The drop in coma emission is more than an order of magnitude and is seen in the OI, OH, CN, NH, and $NH_2$ filters. Over the course of our observations, the morphology of the emission in the different filters does not change noticeably (Fig. 2). After May, the OH and CN emission levels are at the level expected from photoprocesses alone (Fig 4). If the electron density evolved with our scaling law (linear with the gas production and inverse quadratically with the heliocentric distance), we would expect 1000 times more OH emission than was observed (Fig. 5d). Similarly, the predicted [OI] emission from dissociative electron impact of $H_2O$ and $CO_2$ would produce 10x more emission than was observed in July. However, production rates derived from [OI] and NH surface brightnesses assuming photodissociation still require unrealistically high production rates (Fig. 4), and the morphology in the CN filter still resembles that seen in the OI filter, suggesting that collisions with electrons still play a role in the emission observed after May 2015. We propose that the drop between March and June 2015 was thus caused by a change in the number or temperature of projectile electrons available, and that this affects the emission in the different filter in different ways depending on the energy dependence of the relevant electron impact dissociation processes.

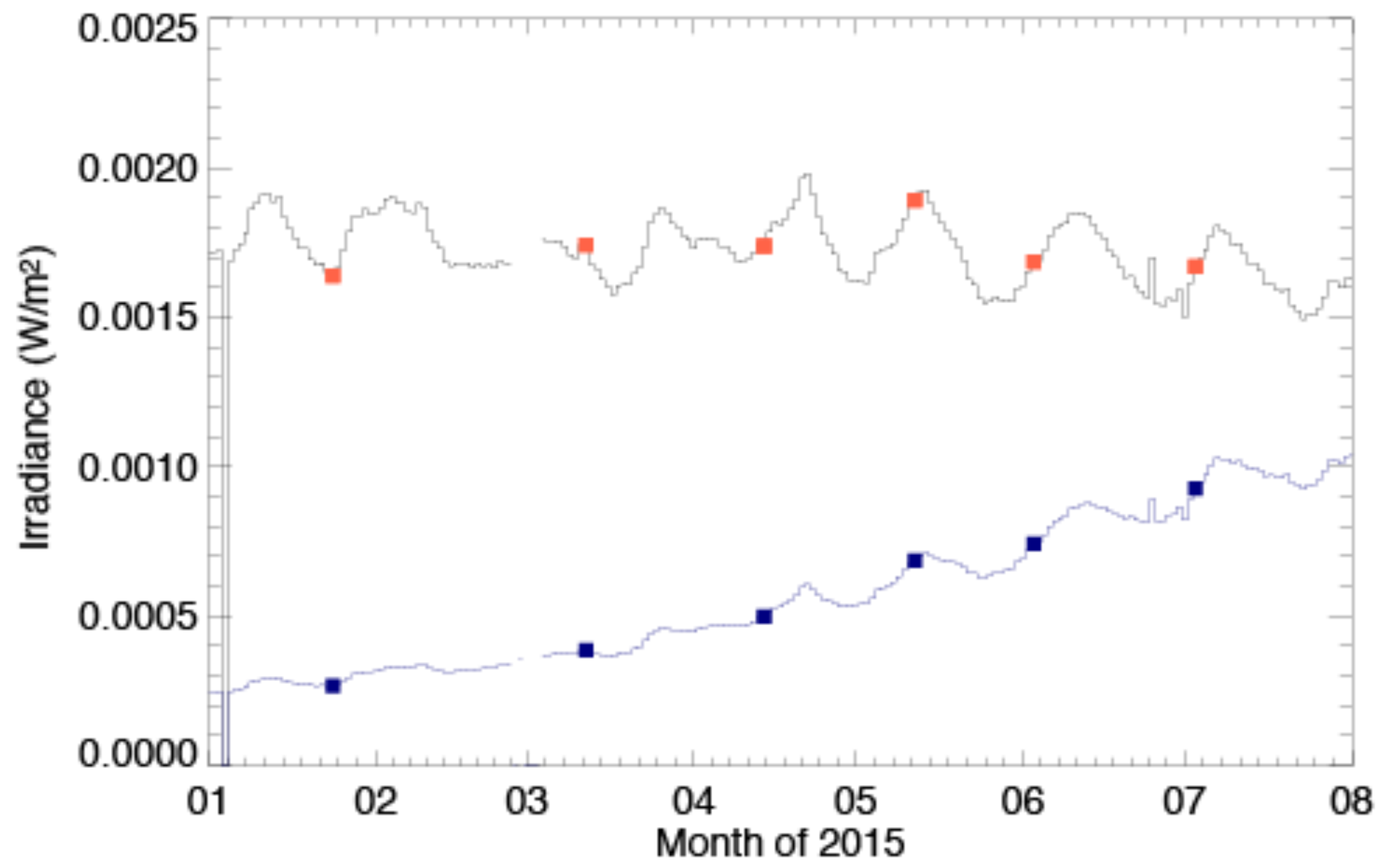

**FIG. 6A** – SDO/EVE integrated irradiances between 30 – 100 nm at 1 AU (black line) and at the heliocentric distance of 67P (blue line). Red and blue dots indicate the times of observations discussed in this paper.

To explain the decrease of emission, we first consider the production of electrons. The *Rosetta*/RPC observations in February, 2015 indicate that the main source of electrons within 256 km of the nucleus is the neutral gas in the coma (Edberg et al. 2015). The far and extreme ultraviolet (FUV/EUV) solar flux determines the photoionization rates of $H_2O$ and other molecules in the coma, and thus also controls the number of electrons available. Inspection of daily averaged solar UV spectra acquired with Extreme Ultraviolet Variability Experiment on board the Solar Dynamics Observatory (SDO/EVE; Woods et al. 2012). We integrated the irradiance spectra over the range 30 – 100 nm[1], the wavelengths that are most efficient in producing electrons by ionizing water (Huebner et al. 1992; Budzien et al. 1994). The results are shown in Fig. 6a. Over the course of our observations, the UV irradiance decreased gradually, and it varied by 10% at time scales of 1 or 2 weeks. Comparing the production rates and UV irradiances in Fig. 6a suggests there is no clear correlation between the short-term solar variations. In this period the comet approached the Sun from 2.5 AU to 1.3 AU, an effect much larger than the weekly UV variations.

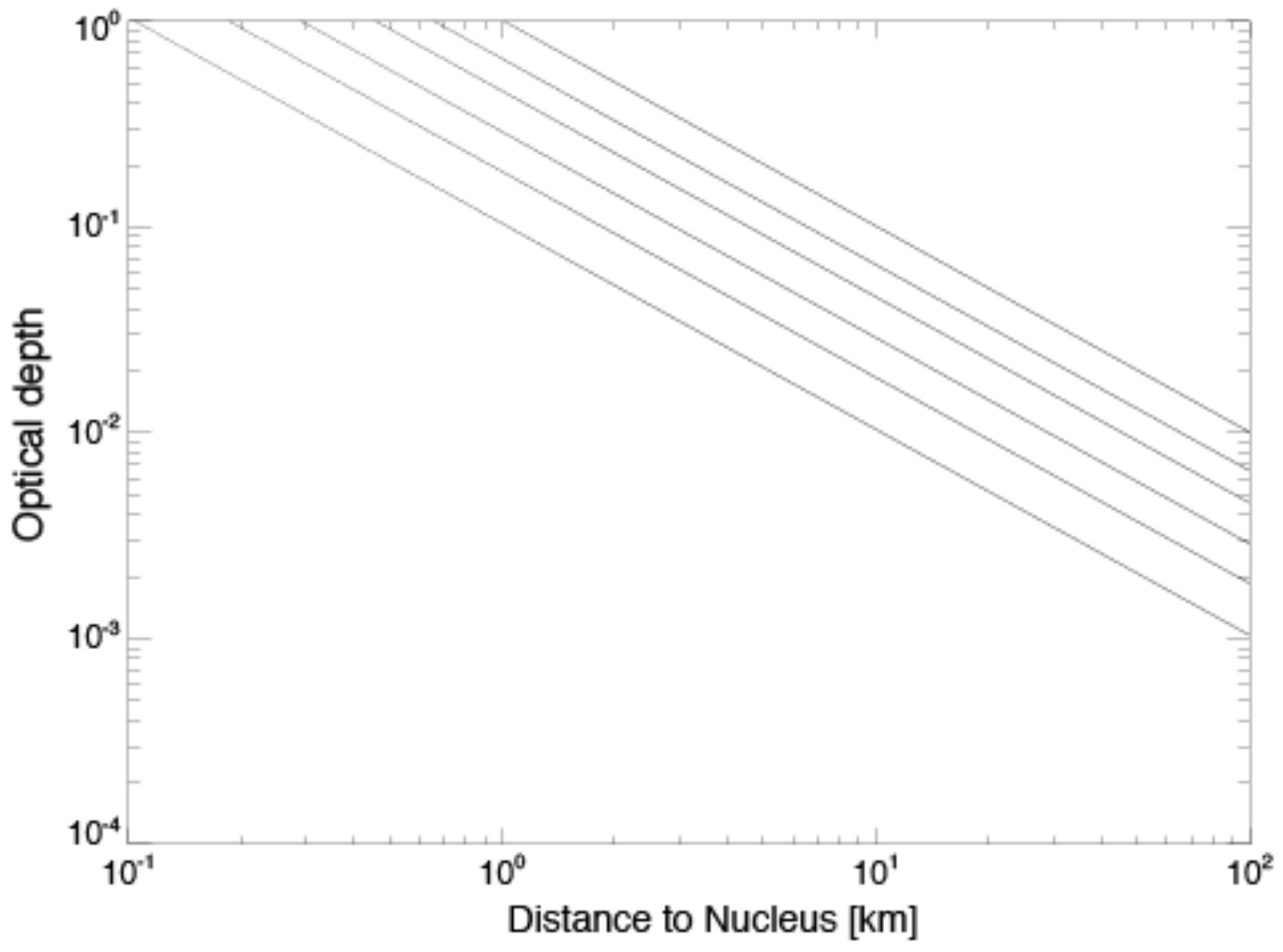

**FIG. 6B** – Optical depth for light with a wavelength of 30 nm, calculated for the different observing dates, January to July 2015 from bottom to top.

With increasing gas production rates, the optical depth of the inner coma increases and fewer photoelectrons are produced. To investigate the optical depth of the coma, we calculated $H_2O$ column densities using a Haser distribution (Appendix A), and assumed water production rates from the empirical formula by Fougere et al. (2016; Fig. 4). The dominant photons for ionization have wavelengths between 30 – 85 nm (Huebner et al. 1992; Budzien et al. 1994). At these wavelengths, $H_2O$ has photo-absorption cross sections of $1.0 \times 10^{-21}$ m$^2$ and $1.6 \times 10^{-21}$ m$^2$, respectively (Phillips et al. 1977). The results are shown in Fig. 6b. The relevant region here is between 1 and 10 km from the surface. In January, only 1 to 10% of photons were lost within this region. This increased to 10 to 64% in July.

[1] We used SDO/EVE level 3 version 5 data acquired with the MEGS-B instrument, available online at http://lasp.colorado.edu/eve/data_access/evewebdata/products/level3/

The electron production thus likely decreased in the region seen by OSIRIS, but not enough to explain the observed drop in emission.

Along with the gas production rates, dust production rates also increased significantly over the course of our observations. Based on *Cassini* observations, it has been suggested that nano-grain charging could result in significant electron depletion (Vigren et al. 2015; Nilsson, et al. 2015b). Reaction rates for attachment of electrons to grains are very low for suprathermal electrons and would mostly affect electrons with much lower energies than those responsible for the electron impact dissociation (>10 eV). We do not expect electron-dust interactions to affect the observed emission.

The most likely explanation of the decrease in emission from the inner coma is a significant decrease in the electron temperature, which reduces reaction rates. At energies of a few eV, when temperatures fall below the appearance thresholds of the dissociative reactions, the excitation of $H_2O$ molecules becomes an effective cooling process (Cravens & Korosmezey 1986). Electrons gain energy by photo-ionization processes, comet-solar wind interactions (Clark et al. 2015), and lose energy through collisional processes, for example (Wegmann et al. 1999):

$$\begin{aligned} H_2O + h\nu &\rightarrow H_2O^+ + e \ [+12.3\ eV] \\ H_2O + e &\rightarrow H_2 + O(\,^1D) + e \ [-12.3\ eV] \\ H_2O + e &\rightarrow OH + H + e \ [-5.1\ eV] \end{aligned} \quad [6]$$

The *Giotto* probe observed maximum ion density 12000 km from the nucleus of 1P/Halley (Häberli et al. 1995). This was attributed to a steep increase of the electron temperature with distance to the nucleus, in turn caused by decreased cooling through collisions between electrons and neutrals.

In the period January – July 2015, the neutral gas density increased much faster ($\sim r_h^{-4.2}$) than the increase of solar radiation ($\sim r_h^{-2}$), which should lead to a decrease in the electron temperature. The electron impact dissociation processes that produce the emission observed with OSIRIS all have relatively high appearance energies (Avakyan et al. 1998). Cooling of the electrons below those energies could lead to an abrupt decrease in the emission.

Lastly, the observed decrease in the emission might be related to changes in the interaction between comet and solar wind. Solar wind electrons have higher temperatures than photo-electrons and its interaction with the coma may heat electrons (Häberli et al. 1996; Clark et al. 2015). Initially, the solar wind could penetrate deep into the coma (Nillson et al. 2015a,b). With increasing production rate plasma boundaries separated the solar wind from the cometary gas (Rubin et al. 2015). By the end of February 2015, the RPC/Ion Composition Analyzer measured that the proton energy spectrum started to change, indicative of an increasing interaction between comet and solar wind.

## SUMMARY AND CONCLUSION

This study uses data from *Rosetta*'s OSIRIS camera system and presents observations of the inner coma of 67P/Churyumov-Gerasimenko acquired with narrow band filters centered on the emission features of OH, OI, CN, NH, and $NH_2$. The observations explore a new regime in cometary science: the inner coma of a low-activity comet at large heliocentric distances, and the *Rosetta* mission allows us to study how this environment changed over time. We developed an extensive image reduction procedure and applied this to observations acquired between January and July, 2015, when the comet's distance to the Sun decreased from 2.5 to 1.3 AU.

The observations in all filters at all epochs indicated surface brightnesses that are one or more orders of magnitude higher than can be explained by photodissociation. Instead, the emission is likely the result of dissociative electron impact excitation, and/or different species. The OH emission can be attributed to electron dissociative excitation of $H_2O$, and the observed intensities are generally consistent with neutral column densities and electron densities observed by other instruments on board *Rosetta*. We attribute most of the [OI] 630 nm emission in the inner coma to electron impact dissociative excitation of $CO_2$, which has a much larger cross section than $H_2O$. The emission detected in the NH filter is likely the result of electron dissociative excitation that produces $OH^+$ ions from $H_2O$ molecules directly in an excited state. The surface brightness levels observed in the CN filter and the similarity between the morphology observed in the CN and [OI] filters suggests that the emission is the product of both electron impact dissociative excitation produced $CO_2^+$ ions and of fluorescent emission by CN radicals. Follow-up studies of the correlation between the emission seen in CN vs. [OI], and OH vs. NH might provide very interesting windows on the different reaction channels from impact on the same species.

The intensity of the emission in the inner coma decreased between March and June, despite increasing gas production rates as the comet approached the Sun. The most likely explanation is that because the gas production rates increased much faster than the ionizing solar radiation, collisions between electrons and neutral water molecules lowered electron temperatures in the inner coma below the activation threshold of the dissociative impact excitation reactions. The increase of optical depth and deflection of the solar wind may have further contributed to the electron cooling.

Our results thus show that the narrowband filters on *Rosetta*/OSIRIS can be used to remotely study the interaction between electrons and the neutral gas in the inner coma. However, the lack of experimental cross sections hampers the interpretation of our results; most urgently needed are those for the production of OI $^1D$ from $CO_2$ and $H_2O$, and of CN from HCN). In addition, we have identified the need for models that combine the physical and chemical processes of cometary gases with plasma characteristics such as electron temperatures.

Impact excitation may matter in different planetary environments through different interaction mechanisms (photo-electrons, solar wind protons and/or electrons). First, since the resulting emission traces the distribution of a parent species, electron impact dissociation could explain the excitation of fragments whenever a steep, radial gradient is seen in the emission morphology of the fragment species, such as in the large-scale jet-like structures around comets (A'Hearn et al. 1986). Second, under conditions similar to that of 67P, dissociative electron impact excitation can lead to significant emission by fragment species. This might lead to detectable emission from Main Belt Comets or even Ceres – or to an overestimate of production rates if only photo-processes are assumed to drive the emission of fragment species.

Comet 67P reached its perihelion on August 13, 2015, at a heliocentric distance of 1.24 AU. The Rosetta spacecraft will continue to orbit its nucleus and study how its coma evolves while the comet moves away from the Sun again.

**Acknowledgements**: The authors wish to thank Mike Combi, Mike Mumma, and the *Rosetta*/Alice team for helpful discussions regarding electron impact processes in the inner coma. Rosetta is an ESA mission with contributions from its member states and NASA. OSIRIS was built by a consortium of the Max-Planck-Institut für Sonnensystemforschung, Göttingen, Germany, the CISAS, University of Padova, Italy, the Laboratoire d'Astrophysique de Marseille, France, the Instituto de Astrofísica de Andalucia, CSIC, Granada, Spain, the Research and Scientific Support Department of the European Space Agency, Noordwijk, The Netherlands, the Instituto Nacional de Técnica Aeroespacial, Madrid, Spain, the Universidad Politéchnica de Madrid, Spain, the Department of Physics and Astronomy of Uppsala University, Sweden, and the Institut für Datentechnik und Kommunikationsnetze der Technischen Universität Braunschweig, Germany. The support of the national funding agencies of Germany (DLR), France (CNES), Italy (ASI), Spain (MEC), Sweden (SNSB), and the ESA Technical Directorate is acknowledged. This work was also supported by NASA JPL contract 1267923 to the University of Maryland (M.F.A'H. and D.B.). M.F.A'H. is also a Gauss Professor of the Akademie der Wissenschaften zu Göttingen and Max-Planck-Institut für Sonnensystemforschung (Germany). We gratefully acknowledge use of SDO/EVE data.

## APPENDIX A: EMISSION MODEL

To aid in our interpretation of our results, we have developed a basic coma model, summarized below. A true model would combine a realistic distribution of the neutral gas, the cometary and solar wind plasma, and all relevant chemical and physical reactions, and the complexity of such a model is outside the scope of this paper.

### A1. Neutral density model

Total water production rates of 67P were estimated using the empirical relation derived between water production *Q($H_2O$)* and heliocentric distance $r_h$ by Fougere et al. (2016):

$$Q(H_2O) \sim 1.02 \times 10^{28}\, r_h^{-4.2} \quad \text{(molecules/s)} \qquad [A1]$$

We assumed constant abundances of $CO_2/H_2O$ = 3%, $CO/H_2O$ = 3%, $O_2/H_2O$ = 3.8%, and $HCN/H_2O$ = 0.1% (LeRoy et al 2015; Fougere et al. 2016; Bieler et al. 2015). For the distribution of fragment species we assumed a standard Haser model (Festou 1981, Combi et al. 2004). The photodissociation rates of the different gases are listed in Table A1, and those were scaled with the heliocentric distance as $1/r_h^2$. We assumed that all gases are accelerated to a bulk velocity $v_g$ within a distance $d$ = 200 km from the center of the nucleus (Combi et al. 2004) using the empirical relation:

$$v_g = 0.85 * r_h^{-1/2} * \left[1 - \exp\left(\frac{-d}{50\ km}\right)\right] \quad \text{(km/s)} \qquad [A2]$$

We assumed the same bulk outflow gas velocity for both parent and fragment species.

### A2. Electrons

The electron density in the coma was measured by the Langmuir (LAP) and mutual impedance probes (MIP) on *Rosetta* between 2015, February 4 – 28 (2.3 AU from the Sun; Edberg et al. 2015). For the electron density we assumed that the number density and radial distribution measured by LAP/MIP scaled linearly with the water production rate and the photoionization rate (thus the square of the heliocentric distance):

$$n_e = \frac{Q_{H2O}}{10^{26}} * \frac{10^{14}}{d} * \frac{3}{r_h^2} \quad (m^{-3}) \qquad [A3]$$

where $n_e$ denotes the electron density and $d$ the distance from the comet's surface. Rather than assuming a temperature distribution, we assumed a fixed electron temperature of 30 eV (3.3 x $10^6$ m/s) for electron fluxes and for the cross sections of electron impact reactions. To calculate the emission resulting from dissociative electron impact dissociation, we multiplied the neutral densities in the coma *n(d)* at a distance $d$ from the nucleus with the local electron density $n_e(d)$, electron velocity $v_e$, and emission cross section σ or with the reaction rate where applicable (Table A1):

$$S(d) = \frac{n(d) * n_e(d) * v_e * \sigma}{4\pi} \quad (ph/s/m^3) \qquad [A4]$$

This is then integrated over the line of sight (from the position of the spacecraft Δ to infinity) to produce the surface brightnesses $B(\theta)$ as a function of the angle between spacecraft-comet line along the comet-Sun line:

$$B(\theta) = \int_{\Delta}^{\infty} \frac{S(d)}{4\pi} \delta d \qquad \text{(ph/s/m}^2\text{/sr)} \qquad [A5]$$

The radial distance *d* from the nucleus in any position in the comet-sun-spacecraft plane can be calculated using the cosine rule.

### A3. OI quenching and transport

The [OI] 630 nm emission is different from the other emission features because it is a forbidden line, and we expect quenching of the OI $^1D$ state by the reaction OI $^1D$ + $H_2O$ → 2 OH to decrease the [OI] 630 nm emission with increasing production rates. The OI $^1D$ state has a long lifetime of 101 s (Atkinson et al. 1997). To assess the effect of the deactivation of OI $^1D$ through collisions with $H_2O$ on the surface brightness profile we combined the Haser model from Sec. A1 and adopted experimental rate coefficients of 2 x $10^{-19}$ $m^3$/s/molecule (Streit et al. 1976). To correct for excited OI atoms lost through collisional quenching, we calculate an effective density *n'(d)* by weighting the neutral gas density function n(d) from equation A3 by a correction factor. Assuming a quenching rate R (Table A1), the distance *l(d)* an OI atom moving with velocity $v_d$ can travel before colliding is:

$$l(d) = \frac{v_d}{R * n(d)} \qquad [A6]$$

Now the fraction of atoms that can emit and thus is not quenched is given by:

$$n'(d) = n(d) * \left[1 - \exp\left(\frac{-l(d)}{v_d * \tau}\right)\right] \qquad [A7]$$

where $\tau$ is the lifetime of the OI $^1D$ state (101 s), and *n(d)* the density of neutral water at a distance d of the surface. The effect of quenching on the surface brightness is shown in Fig. A1. We assumed a fixed distance between comet and spacecraft of 100 km to evaluate the effect of quenching and have added the [OI] emission from all $H_2O$ and $CO_2$ processes (the dominant contributions to the OI emission, see Fig. 5 in the main text). In January, only the first kilometer is affected by quenching in the inner coma. In July, the inner 3 kilometers are affected, and at 1 km approximately the observed surface brightness is approximately 50% of the emitted light.

In addition, because of the long lifetimes of the $^1D$ state, only part of the atoms decay to the ground state within the field of view. We account for this with a second correction factor:

$$n''(d) = n'(d) * \left[1 - \exp\left(\frac{-d}{v_d * \tau}\right)\right] \qquad [A8]$$

The results, again for a fixed spacecraft distance of 100 km, are shown in Figs. 5c and 5d. The effect of the OI $^1D$ lifetime is dramatic and flattens the surface emission in the first few kilometers around the nucleus.

### A4. Optical depth

With increasing gas production rates, the optical depth of the inner coma increases and fewer photoelectrons are produced. To investigate the optical depth of the coma, we calculated $H_2O$ column densities using a Haser distribution (Sec. 3.3), and assumed water production rates from the empirical formula by Fougere et al. (2016; see also Fig. 4a). The dominant photons for ionization have wavelengths between 30 – 85 nm (Huebner et al. 1992; Budzien et al. 1994). At these wavelengths, $H_2O$ has photo-absorption cross sections of 1.0 x $10^{-21}$ $m^2$ and 1.6 x $10^{-21}$ $m^2$, respectively (Phillips et al. 1977). The results are shown in Fig. 6b. The relevant region here is between 1 and 10 km from the surface. In January, only 1 to 10% of photons were lost within this region. The effects of the increasing optical depth on the neutrals and electrons in the inner coma is not included in the surface brightness model.

Table A1. Assumed reaction rates (at $r_h$ = 1 AU) and cross sections

| Reaction | Product | Cross section or rate | Reference |
|---|---|---|---|
| $H_2O + h\nu$ | any | $1.04 \times 10^{-5}$ s$^{-1}$ | Combi et al. 2004 |
| $H_2O + h\nu$ | OH + $h\nu$ | $8.5 \times 10^{-4}$ s$^{-1}$ | Combi et al. 2004 |
| $H_2O + h\nu$ | OI ($^1$S + $^1$D) | $8.6 \times 10^{-7}$ s$^{-1}$ | Bhardwaj & Raghuram 2012 |
| $CO_2 + h\nu$ | any | $2.2 \times 10^{-6}$ s$^{-1}$ | Weaver et al. 1999 |
| $CO_2 + h\nu$ | OI ($^1$D) | $1.9 \times 10^{-6}$ s$^{-1}$ | Bhardwaj & Raghuram 2012 |
| CO + $h\nu$ | any | $3.3 \times 10^{-6}$ s$^{-1}$ | Huebner et al. 1992 |
| CO + $h\nu$ | OI ($^1$D) | $9.1 \times 10^{-8}$ s$^{-1}$ | Bhardwaj & Raghuram 2012 |
| $O_2 + h\nu$ | any | $4.5 \times 10^{-6}$ s$^{-1}$ | Huebner et al. 1992 |
| $O_2 + h\nu$ | OI ($^1$D) | $4.0 \times 10^{-6}$ s$^{-1}$ | Huebner et al. 1992 |
| $NH_3 + h\nu$ | any | | |
| $NH_3 + h\nu$ | NH | | |
| HCN + $h\nu$ | any | $7.5 \times 10^{-5}$ s$^{-1}$ | Huebner et al. 1992 |
| HCN + $h\nu$ | CN | $7.5 \times 10^{-5}$ s$^{-1}$ | Huebner et al. 1992 |
| HCN + $h\nu$ | CN ($B^2\Sigma^+$) | $4.5 \times 10^{-5}$ s$^{-1}$ | Fray et al. 2005 |
| $H_2O$ + e- | OH ($A^2\Sigma^+$) | $8.5 \times 10^{-22}$ m$^2$ | Avakyan et al. 1998 |
| $H_2O$ + e- | OI ($^1$S + $^1$D) | $9.7 \times 10^{-16}$ m$^3$s$^{-1}$ | Bhardwaj & Raghuram 2012 |
| $CO_2 + e^-$ | OI ($^1$S + $^1$D) | $5.3 \times 10^{-14}$ m$^3$s$^{-1}$ | Bhardwaj & Raghuram 2012 |
| CO + e- | OI ($^1$S + $^1$D) | $2.9 \times 10^{-16}$ m$^3$s$^{-1}$ | Bhardwaj & Raghuram 2012 |
| $O_2$ + e- | OI ($^1$S + $^1$D) | $1 \times 10^{-14}$ m$^3$s$^{-1}$ | Estimated from Avakyan et al. 1998 |
| $O_2$ + e- | $X^{q+}$ | $1.5 \times 10^{-20}$ m$^2$ | Straub et al. 1996 |
| $NH_3$ + e- | NH ($A^3\Pi_1$) | $2.8 \times 10^{-22}$ m$^2$ | Müller & Schulz 1992 |
| OI $^1$D + $H_2O$ | 2OH | $2 \times 10^{-16}$ m$^3$/s/molecule | Streit et al. 1976 |